\newif\ifarxiv
\arxivtrue

\documentclass[journal,a4paper]{IEEEtran}
\usepackage{amsthm}
\usepackage{thm-restate}
\usepackage{cite}
\usepackage{subcaption}
\usepackage[cmex10]{amsmath}
\usepackage{amssymb,amsfonts}
\usepackage{algorithmic}
\usepackage{graphicx}
\usepackage{textcomp}
\usepackage{xcolor}
\usepackage{tabularray}
\usepackage{contour}
\usepackage{mathtools}
\def\BibTeX{{\rm B\kern-.05em{\sc i\kern-.025em b}\kern-.08em
    T\kern-.1667em\lower.7ex\hbox{E}\kern-.125em}}
\usepackage{eqnarray}
\usepackage{hyperref}

\usepackage[utf8]{inputenc}
\usepackage{mleftright}       %
\mleftright                   %

\usepackage{pgfplots}
 \usetikzlibrary{arrows.meta}
 \usetikzlibrary{backgrounds}
 \usepgfplotslibrary{patchplots}
 \usepgfplotslibrary{fillbetween}
 \pgfplotsset{%
      compat=1.18,
     layers/standard/.define layer set={%
         background,axis background,axis grid,axis ticks,axis lines,axis tick labels,pre main,main,axis descriptions,axis foreground%
     }{
         grid style={/pgfplots/on layer=axis grid},%
         tick style={/pgfplots/on layer=axis ticks},%
         axis line style={/pgfplots/on layer=axis lines},%
         label style={/pgfplots/on layer=axis descriptions},%
         legend style={/pgfplots/on layer=axis descriptions},%
         title style={/pgfplots/on layer=axis descriptions},%
         colorbar style={/pgfplots/on layer=axis descriptions},%
         ticklabel style={/pgfplots/on layer=axis tick labels},%
         axis background@ style={/pgfplots/on layer=axis background},%
         3d box foreground style={/pgfplots/on layer=axis foreground},%
     },
 }

 \pgfplotsset{
 colormap={plots1}{rgb(0.00000000)=(0.30980000,0.10090000,0.23840000)
 rgb(0.00392157)=(0.30690000,0.10310000,0.24320000)
 rgb(0.00784314)=(0.30410000,0.10530000,0.24800000)
 rgb(0.01176471)=(0.30120000,0.10760000,0.25300000)
 rgb(0.01568627)=(0.29840000,0.11020000,0.25800000)
 rgb(0.01960784)=(0.29550000,0.11280000,0.26310000)
 rgb(0.02352941)=(0.29270000,0.11540000,0.26820000)
 rgb(0.02745098)=(0.28980000,0.11830000,0.27350000)
 rgb(0.03137255)=(0.28690000,0.12120000,0.27890000)
 rgb(0.03529412)=(0.28410000,0.12430000,0.28430000)
 rgb(0.03921569)=(0.28120000,0.12750000,0.28990000)
 rgb(0.04313725)=(0.27830000,0.13090000,0.29550000)
 rgb(0.04705882)=(0.27540000,0.13430000,0.30120000)
 rgb(0.05098039)=(0.27250000,0.13790000,0.30700000)
 rgb(0.05490196)=(0.26960000,0.14160000,0.31290000)
 rgb(0.05882353)=(0.26670000,0.14550000,0.31890000)
 rgb(0.06274510)=(0.26370000,0.14940000,0.32490000)
 rgb(0.06666667)=(0.26080000,0.15350000,0.33110000)
 rgb(0.07058824)=(0.25780000,0.15770000,0.33730000)
 rgb(0.07450980)=(0.25490000,0.16210000,0.34360000)
 rgb(0.07843137)=(0.25190000,0.16650000,0.34990000)
 rgb(0.08235294)=(0.24900000,0.17110000,0.35630000)
 rgb(0.08627451)=(0.24600000,0.17590000,0.36280000)
 rgb(0.09019608)=(0.24310000,0.18070000,0.36930000)
 rgb(0.09411765)=(0.24020000,0.18580000,0.37590000)
 rgb(0.09803922)=(0.23730000,0.19080000,0.38250000)
 rgb(0.10196078)=(0.23440000,0.19610000,0.38920000)
 rgb(0.10588235)=(0.23160000,0.20140000,0.39590000)
 rgb(0.10980392)=(0.22870000,0.20690000,0.40260000)
 rgb(0.11372549)=(0.22600000,0.21250000,0.40940000)
 rgb(0.11764706)=(0.22330000,0.21820000,0.41610000)
 rgb(0.12156863)=(0.22070000,0.22410000,0.42290000)
 rgb(0.12549020)=(0.21820000,0.23000000,0.42970000)
 rgb(0.12941176)=(0.21580000,0.23610000,0.43650000)
 rgb(0.13333333)=(0.21340000,0.24220000,0.44330000)
 rgb(0.13725490)=(0.21130000,0.24850000,0.45000000)
 rgb(0.14117647)=(0.20920000,0.25480000,0.45680000)
 rgb(0.14509804)=(0.20740000,0.26130000,0.46350000)
 rgb(0.14901961)=(0.20570000,0.26780000,0.47030000)
 rgb(0.15294118)=(0.20430000,0.27440000,0.47700000)
 rgb(0.15686275)=(0.20300000,0.28110000,0.48360000)
 rgb(0.16078431)=(0.20200000,0.28790000,0.49020000)
 rgb(0.16470588)=(0.20130000,0.29480000,0.49680000)
 rgb(0.16862745)=(0.20080000,0.30170000,0.50340000)
 rgb(0.17254902)=(0.20070000,0.30870000,0.50990000)
 rgb(0.17647059)=(0.20080000,0.31580000,0.51640000)
 rgb(0.18039216)=(0.20130000,0.32290000,0.52280000)
 rgb(0.18431373)=(0.20210000,0.33010000,0.52920000)
 rgb(0.18823529)=(0.20330000,0.33740000,0.53550000)
 rgb(0.19215686)=(0.20480000,0.34470000,0.54180000)
 rgb(0.19607843)=(0.20680000,0.35200000,0.54800000)
 rgb(0.20000000)=(0.20900000,0.35940000,0.55420000)
 rgb(0.20392157)=(0.21170000,0.36690000,0.56030000)
 rgb(0.20784314)=(0.21470000,0.37430000,0.56640000)
 rgb(0.21176471)=(0.21810000,0.38190000,0.57250000)
 rgb(0.21568627)=(0.22180000,0.38940000,0.57850000)
 rgb(0.21960784)=(0.22590000,0.39700000,0.58450000)
 rgb(0.22352941)=(0.23040000,0.40460000,0.59040000)
 rgb(0.22745098)=(0.23530000,0.41230000,0.59620000)
 rgb(0.23137255)=(0.24040000,0.42000000,0.60210000)
 rgb(0.23529412)=(0.24590000,0.42770000,0.60790000)
 rgb(0.23921569)=(0.25170000,0.43540000,0.61360000)
 rgb(0.24313725)=(0.25780000,0.44320000,0.61930000)
 rgb(0.24705882)=(0.26420000,0.45100000,0.62500000)
 rgb(0.25098039)=(0.27090000,0.45870000,0.63060000)
 rgb(0.25490196)=(0.27790000,0.46650000,0.63620000)
 rgb(0.25882353)=(0.28510000,0.47440000,0.64170000)
 rgb(0.26274510)=(0.29260000,0.48220000,0.64720000)
 rgb(0.26666667)=(0.30030000,0.49000000,0.65270000)
 rgb(0.27058824)=(0.30830000,0.49780000,0.65810000)
 rgb(0.27450980)=(0.31650000,0.50560000,0.66350000)
 rgb(0.27843137)=(0.32480000,0.51350000,0.66880000)
 rgb(0.28235294)=(0.33340000,0.52130000,0.67410000)
 rgb(0.28627451)=(0.34220000,0.52910000,0.67930000)
 rgb(0.29019608)=(0.35120000,0.53690000,0.68440000)
 rgb(0.29411765)=(0.36030000,0.54460000,0.68950000)
 rgb(0.29803922)=(0.36960000,0.55240000,0.69460000)
 rgb(0.30196078)=(0.37900000,0.56010000,0.69960000)
 rgb(0.30588235)=(0.38860000,0.56780000,0.70450000)
 rgb(0.30980392)=(0.39830000,0.57540000,0.70930000)
 rgb(0.31372549)=(0.40820000,0.58300000,0.71410000)
 rgb(0.31764706)=(0.41820000,0.59050000,0.71870000)
 rgb(0.32156863)=(0.42830000,0.59800000,0.72330000)
 rgb(0.32549020)=(0.43850000,0.60550000,0.72780000)
 rgb(0.32941176)=(0.44890000,0.61280000,0.73220000)
 rgb(0.33333333)=(0.45930000,0.62010000,0.73640000)
 rgb(0.33725490)=(0.46980000,0.62730000,0.74060000)
 rgb(0.34117647)=(0.48040000,0.63450000,0.74460000)
 rgb(0.34509804)=(0.49100000,0.64150000,0.74850000)
 rgb(0.34901961)=(0.50170000,0.64840000,0.75230000)
 rgb(0.35294118)=(0.51250000,0.65520000,0.75590000)
 rgb(0.35686275)=(0.52330000,0.66200000,0.75930000)
 rgb(0.36078431)=(0.53410000,0.66850000,0.76260000)
 rgb(0.36470588)=(0.54490000,0.67490000,0.76560000)
 rgb(0.36862745)=(0.55580000,0.68120000,0.76850000)
 rgb(0.37254902)=(0.56660000,0.68740000,0.77120000)
 rgb(0.37647059)=(0.57750000,0.69330000,0.77370000)
 rgb(0.38039216)=(0.58830000,0.69910000,0.77600000)
 rgb(0.38431373)=(0.59900000,0.70470000,0.77800000)
 rgb(0.38823529)=(0.60980000,0.71010000,0.77980000)
 rgb(0.39215686)=(0.62040000,0.71530000,0.78140000)
 rgb(0.39607843)=(0.63100000,0.72030000,0.78260000)
 rgb(0.40000000)=(0.64140000,0.72500000,0.78360000)
 rgb(0.40392157)=(0.65180000,0.72950000,0.78440000)
 rgb(0.40784314)=(0.66200000,0.73380000,0.78480000)
 rgb(0.41176471)=(0.67210000,0.73780000,0.78500000)
 rgb(0.41568627)=(0.68200000,0.74150000,0.78480000)
 rgb(0.41960784)=(0.69170000,0.74490000,0.78430000)
 rgb(0.42352941)=(0.70130000,0.74810000,0.78360000)
 rgb(0.42745098)=(0.71060000,0.75100000,0.78250000)
 rgb(0.43137255)=(0.71980000,0.75360000,0.78100000)
 rgb(0.43529412)=(0.72870000,0.75580000,0.77930000)
 rgb(0.43921569)=(0.73730000,0.75780000,0.77720000)
 rgb(0.44313725)=(0.74570000,0.75940000,0.77480000)
 rgb(0.44705882)=(0.75380000,0.76080000,0.77200000)
 rgb(0.45098039)=(0.76160000,0.76180000,0.76900000)
 rgb(0.45490196)=(0.76910000,0.76250000,0.76560000)
 rgb(0.45882353)=(0.77640000,0.76280000,0.76190000)
 rgb(0.46274510)=(0.78320000,0.76290000,0.75780000)
 rgb(0.46666667)=(0.78980000,0.76260000,0.75350000)
 rgb(0.47058824)=(0.79610000,0.76200000,0.74880000)
 rgb(0.47450980)=(0.80200000,0.76110000,0.74390000)
 rgb(0.47843137)=(0.80760000,0.75980000,0.73870000)
 rgb(0.48235294)=(0.81280000,0.75830000,0.73320000)
 rgb(0.48627451)=(0.81770000,0.75650000,0.72740000)
 rgb(0.49019608)=(0.82220000,0.75430000,0.72130000)
 rgb(0.49411765)=(0.82640000,0.75190000,0.71510000)
 rgb(0.49803922)=(0.83030000,0.74920000,0.70860000)
 rgb(0.50196078)=(0.83390000,0.74630000,0.70180000)
 rgb(0.50588235)=(0.83710000,0.74310000,0.69490000)
 rgb(0.50980392)=(0.84000000,0.73960000,0.68770000)
 rgb(0.51372549)=(0.84250000,0.73590000,0.68040000)
 rgb(0.51764706)=(0.84480000,0.73200000,0.67280000)
 rgb(0.52156863)=(0.84670000,0.72780000,0.66520000)
 rgb(0.52549020)=(0.84840000,0.72340000,0.65730000)
 rgb(0.52941176)=(0.84970000,0.71890000,0.64930000)
 rgb(0.53333333)=(0.85080000,0.71410000,0.64120000)
 rgb(0.53725490)=(0.85160000,0.70920000,0.63300000)
 rgb(0.54117647)=(0.85220000,0.70410000,0.62460000)
 rgb(0.54509804)=(0.85250000,0.69880000,0.61620000)
 rgb(0.54901961)=(0.85260000,0.69340000,0.60760000)
 rgb(0.55294118)=(0.85240000,0.68780000,0.59900000)
 rgb(0.55686275)=(0.85200000,0.68210000,0.59030000)
 rgb(0.56078431)=(0.85140000,0.67620000,0.58150000)
 rgb(0.56470588)=(0.85050000,0.67030000,0.57270000)
 rgb(0.56862745)=(0.84950000,0.66420000,0.56380000)
 rgb(0.57254902)=(0.84830000,0.65800000,0.55490000)
 rgb(0.57647059)=(0.84690000,0.65170000,0.54590000)
 rgb(0.58039216)=(0.84530000,0.64530000,0.53690000)
 rgb(0.58431373)=(0.84350000,0.63880000,0.52780000)
 rgb(0.58823529)=(0.84160000,0.63220000,0.51880000)
 rgb(0.59215686)=(0.83950000,0.62550000,0.50970000)
 rgb(0.59607843)=(0.83720000,0.61880000,0.50050000)
 rgb(0.60000000)=(0.83480000,0.61190000,0.49140000)
 rgb(0.60392157)=(0.83230000,0.60500000,0.48230000)
 rgb(0.60784314)=(0.82960000,0.59800000,0.47310000)
 rgb(0.61176471)=(0.82670000,0.59090000,0.46400000)
 rgb(0.61568627)=(0.82370000,0.58380000,0.45480000)
 rgb(0.61960784)=(0.82060000,0.57650000,0.44570000)
 rgb(0.62352941)=(0.81730000,0.56920000,0.43660000)
 rgb(0.62745098)=(0.81390000,0.56190000,0.42750000)
 rgb(0.63137255)=(0.81040000,0.55450000,0.41830000)
 rgb(0.63529412)=(0.80670000,0.54690000,0.40930000)
 rgb(0.63921569)=(0.80290000,0.53940000,0.40020000)
 rgb(0.64313725)=(0.79890000,0.53170000,0.39110000)
 rgb(0.64705882)=(0.79480000,0.52400000,0.38210000)
 rgb(0.65098039)=(0.79060000,0.51620000,0.37310000)
 rgb(0.65490196)=(0.78620000,0.50840000,0.36420000)
 rgb(0.65882353)=(0.78170000,0.50040000,0.35530000)
 rgb(0.66274510)=(0.77710000,0.49240000,0.34640000)
 rgb(0.66666667)=(0.77230000,0.48440000,0.33760000)
 rgb(0.67058824)=(0.76730000,0.47620000,0.32880000)
 rgb(0.67450980)=(0.76220000,0.46800000,0.32010000)
 rgb(0.67843137)=(0.75700000,0.45970000,0.31140000)
 rgb(0.68235294)=(0.75160000,0.45140000,0.30280000)
 rgb(0.68627451)=(0.74600000,0.44300000,0.29430000)
 rgb(0.69019608)=(0.74030000,0.43450000,0.28590000)
 rgb(0.69411765)=(0.73450000,0.42600000,0.27760000)
 rgb(0.69803922)=(0.72850000,0.41740000,0.26940000)
 rgb(0.70196078)=(0.72230000,0.40880000,0.26130000)
 rgb(0.70588235)=(0.71600000,0.40010000,0.25340000)
 rgb(0.70980392)=(0.70960000,0.39140000,0.24550000)
 rgb(0.71372549)=(0.70300000,0.38260000,0.23780000)
 rgb(0.71764706)=(0.69630000,0.37380000,0.23030000)
 rgb(0.72156863)=(0.68940000,0.36500000,0.22290000)
 rgb(0.72549020)=(0.68250000,0.35620000,0.21570000)
 rgb(0.72941176)=(0.67540000,0.34740000,0.20870000)
 rgb(0.73333333)=(0.66820000,0.33850000,0.20190000)
 rgb(0.73725490)=(0.66090000,0.32970000,0.19530000)
 rgb(0.74117647)=(0.65350000,0.32090000,0.18900000)
 rgb(0.74509804)=(0.64600000,0.31220000,0.18290000)
 rgb(0.74901961)=(0.63850000,0.30350000,0.17700000)
 rgb(0.75294118)=(0.63090000,0.29490000,0.17140000)
 rgb(0.75686275)=(0.62330000,0.28630000,0.16600000)
 rgb(0.76078431)=(0.61560000,0.27790000,0.16100000)
 rgb(0.76470588)=(0.60790000,0.26950000,0.15620000)
 rgb(0.76862745)=(0.60020000,0.26120000,0.15160000)
 rgb(0.77254902)=(0.59250000,0.25310000,0.14750000)
 rgb(0.77647059)=(0.58490000,0.24500000,0.14350000)
 rgb(0.78039216)=(0.57720000,0.23720000,0.13990000)
 rgb(0.78431373)=(0.56960000,0.22950000,0.13660000)
 rgb(0.78823529)=(0.56210000,0.22190000,0.13360000)
 rgb(0.79215686)=(0.55460000,0.21450000,0.13080000)
 rgb(0.79607843)=(0.54720000,0.20730000,0.12840000)
 rgb(0.80000000)=(0.53990000,0.20020000,0.12620000)
 rgb(0.80392157)=(0.53270000,0.19340000,0.12430000)
 rgb(0.80784314)=(0.52560000,0.18680000,0.12270000)
 rgb(0.81176471)=(0.51860000,0.18030000,0.12140000)
 rgb(0.81568627)=(0.51170000,0.17410000,0.12030000)
 rgb(0.81960784)=(0.50500000,0.16810000,0.11950000)
 rgb(0.82352941)=(0.49840000,0.16230000,0.11890000)
 rgb(0.82745098)=(0.49190000,0.15670000,0.11850000)
 rgb(0.83137255)=(0.48560000,0.15130000,0.11840000)
 rgb(0.83529412)=(0.47940000,0.14620000,0.11840000)
 rgb(0.83921569)=(0.47330000,0.14130000,0.11870000)
 rgb(0.84313725)=(0.46740000,0.13650000,0.11910000)
 rgb(0.84705882)=(0.46160000,0.13200000,0.11970000)
 rgb(0.85098039)=(0.45600000,0.12770000,0.12050000)
 rgb(0.85490196)=(0.45050000,0.12360000,0.12150000)
 rgb(0.85882353)=(0.44510000,0.11980000,0.12260000)
 rgb(0.86274510)=(0.43990000,0.11620000,0.12390000)
 rgb(0.86666667)=(0.43480000,0.11280000,0.12540000)
 rgb(0.87058824)=(0.42980000,0.10960000,0.12690000)
 rgb(0.87450980)=(0.42500000,0.10650000,0.12860000)
 rgb(0.87843137)=(0.42030000,0.10370000,0.13050000)
 rgb(0.88235294)=(0.41570000,0.10110000,0.13240000)
 rgb(0.88627451)=(0.41120000,0.09870000,0.13450000)
 rgb(0.89019608)=(0.40680000,0.09640000,0.13660000)
 rgb(0.89411765)=(0.40260000,0.09440000,0.13890000)
 rgb(0.89803922)=(0.39840000,0.09250000,0.14130000)
 rgb(0.90196078)=(0.39430000,0.09090000,0.14380000)
 rgb(0.90588235)=(0.39040000,0.08940000,0.14640000)
 rgb(0.90980392)=(0.38650000,0.08810000,0.14900000)
 rgb(0.91372549)=(0.38270000,0.08700000,0.15180000)
 rgb(0.91764706)=(0.37900000,0.08590000,0.15470000)
 rgb(0.92156863)=(0.37530000,0.08510000,0.15760000)
 rgb(0.92549020)=(0.37170000,0.08450000,0.16060000)
 rgb(0.92941176)=(0.36820000,0.08410000,0.16380000)
 rgb(0.93333333)=(0.36480000,0.08370000,0.16700000)
 rgb(0.93725490)=(0.36140000,0.08350000,0.17030000)
 rgb(0.94117647)=(0.35810000,0.08350000,0.17360000)
 rgb(0.94509804)=(0.35480000,0.08360000,0.17710000)
 rgb(0.94901961)=(0.35160000,0.08390000,0.18060000)
 rgb(0.95294118)=(0.34840000,0.08430000,0.18420000)
 rgb(0.95686275)=(0.34530000,0.08480000,0.18790000)
 rgb(0.96078431)=(0.34220000,0.08540000,0.19170000)
 rgb(0.96470588)=(0.33910000,0.08620000,0.19550000)
 rgb(0.96862745)=(0.33610000,0.08720000,0.19940000)
 rgb(0.97254902)=(0.33310000,0.08820000,0.20350000)
 rgb(0.97647059)=(0.33010000,0.08940000,0.20750000)
 rgb(0.98039216)=(0.32720000,0.09070000,0.21170000)
 rgb(0.98431373)=(0.32420000,0.09210000,0.21600000)
 rgb(0.98823529)=(0.32130000,0.09360000,0.22030000)
 rgb(0.99215686)=(0.31840000,0.09530000,0.22470000)
 rgb(0.99607843)=(0.31550000,0.09700000,0.22920000)
 rgb(1.00000000)=(0.31260000,0.09890000,0.23380000)},
 }

 \pgfplotsset{
 colormap={plots1}{rgb(0.00000000)=(0.30980000,0.10090000,0.23840000)
 rgb(0.00392157)=(0.30690000,0.10310000,0.24320000)
 rgb(0.00784314)=(0.30410000,0.10530000,0.24800000)
 rgb(0.01176471)=(0.30120000,0.10760000,0.25300000)
 rgb(0.01568627)=(0.29840000,0.11020000,0.25800000)
 rgb(0.01960784)=(0.29550000,0.11280000,0.26310000)
 rgb(0.02352941)=(0.29270000,0.11540000,0.26820000)
 rgb(0.02745098)=(0.28980000,0.11830000,0.27350000)
 rgb(0.03137255)=(0.28690000,0.12120000,0.27890000)
 rgb(0.03529412)=(0.28410000,0.12430000,0.28430000)
 rgb(0.03921569)=(0.28120000,0.12750000,0.28990000)
 rgb(0.04313725)=(0.27830000,0.13090000,0.29550000)
 rgb(0.04705882)=(0.27540000,0.13430000,0.30120000)
 rgb(0.05098039)=(0.27250000,0.13790000,0.30700000)
 rgb(0.05490196)=(0.26960000,0.14160000,0.31290000)
 rgb(0.05882353)=(0.26670000,0.14550000,0.31890000)
 rgb(0.06274510)=(0.26370000,0.14940000,0.32490000)
 rgb(0.06666667)=(0.26080000,0.15350000,0.33110000)
 rgb(0.07058824)=(0.25780000,0.15770000,0.33730000)
 rgb(0.07450980)=(0.25490000,0.16210000,0.34360000)
 rgb(0.07843137)=(0.25190000,0.16650000,0.34990000)
 rgb(0.08235294)=(0.24900000,0.17110000,0.35630000)
 rgb(0.08627451)=(0.24600000,0.17590000,0.36280000)
 rgb(0.09019608)=(0.24310000,0.18070000,0.36930000)
 rgb(0.09411765)=(0.24020000,0.18580000,0.37590000)
 rgb(0.09803922)=(0.23730000,0.19080000,0.38250000)
 rgb(0.10196078)=(0.23440000,0.19610000,0.38920000)
 rgb(0.10588235)=(0.23160000,0.20140000,0.39590000)
 rgb(0.10980392)=(0.22870000,0.20690000,0.40260000)
 rgb(0.11372549)=(0.22600000,0.21250000,0.40940000)
 rgb(0.11764706)=(0.22330000,0.21820000,0.41610000)
 rgb(0.12156863)=(0.22070000,0.22410000,0.42290000)
 rgb(0.12549020)=(0.21820000,0.23000000,0.42970000)
 rgb(0.12941176)=(0.21580000,0.23610000,0.43650000)
 rgb(0.13333333)=(0.21340000,0.24220000,0.44330000)
 rgb(0.13725490)=(0.21130000,0.24850000,0.45000000)
 rgb(0.14117647)=(0.20920000,0.25480000,0.45680000)
 rgb(0.14509804)=(0.20740000,0.26130000,0.46350000)
 rgb(0.14901961)=(0.20570000,0.26780000,0.47030000)
 rgb(0.15294118)=(0.20430000,0.27440000,0.47700000)
 rgb(0.15686275)=(0.20300000,0.28110000,0.48360000)
 rgb(0.16078431)=(0.20200000,0.28790000,0.49020000)
 rgb(0.16470588)=(0.20130000,0.29480000,0.49680000)
 rgb(0.16862745)=(0.20080000,0.30170000,0.50340000)
 rgb(0.17254902)=(0.20070000,0.30870000,0.50990000)
 rgb(0.17647059)=(0.20080000,0.31580000,0.51640000)
 rgb(0.18039216)=(0.20130000,0.32290000,0.52280000)
 rgb(0.18431373)=(0.20210000,0.33010000,0.52920000)
 rgb(0.18823529)=(0.20330000,0.33740000,0.53550000)
 rgb(0.19215686)=(0.20480000,0.34470000,0.54180000)
 rgb(0.19607843)=(0.20680000,0.35200000,0.54800000)
 rgb(0.20000000)=(0.20900000,0.35940000,0.55420000)
 rgb(0.20392157)=(0.21170000,0.36690000,0.56030000)
 rgb(0.20784314)=(0.21470000,0.37430000,0.56640000)
 rgb(0.21176471)=(0.21810000,0.38190000,0.57250000)
 rgb(0.21568627)=(0.22180000,0.38940000,0.57850000)
 rgb(0.21960784)=(0.22590000,0.39700000,0.58450000)
 rgb(0.22352941)=(0.23040000,0.40460000,0.59040000)
 rgb(0.22745098)=(0.23530000,0.41230000,0.59620000)
 rgb(0.23137255)=(0.24040000,0.42000000,0.60210000)
 rgb(0.23529412)=(0.24590000,0.42770000,0.60790000)
 rgb(0.23921569)=(0.25170000,0.43540000,0.61360000)
 rgb(0.24313725)=(0.25780000,0.44320000,0.61930000)
 rgb(0.24705882)=(0.26420000,0.45100000,0.62500000)
 rgb(0.25098039)=(0.27090000,0.45870000,0.63060000)
 rgb(0.25490196)=(0.27790000,0.46650000,0.63620000)
 rgb(0.25882353)=(0.28510000,0.47440000,0.64170000)
 rgb(0.26274510)=(0.29260000,0.48220000,0.64720000)
 rgb(0.26666667)=(0.30030000,0.49000000,0.65270000)
 rgb(0.27058824)=(0.30830000,0.49780000,0.65810000)
 rgb(0.27450980)=(0.31650000,0.50560000,0.66350000)
 rgb(0.27843137)=(0.32480000,0.51350000,0.66880000)
 rgb(0.28235294)=(0.33340000,0.52130000,0.67410000)
 rgb(0.28627451)=(0.34220000,0.52910000,0.67930000)
 rgb(0.29019608)=(0.35120000,0.53690000,0.68440000)
 rgb(0.29411765)=(0.36030000,0.54460000,0.68950000)
 rgb(0.29803922)=(0.36960000,0.55240000,0.69460000)
 rgb(0.30196078)=(0.37900000,0.56010000,0.69960000)
 rgb(0.30588235)=(0.38860000,0.56780000,0.70450000)
 rgb(0.30980392)=(0.39830000,0.57540000,0.70930000)
 rgb(0.31372549)=(0.40820000,0.58300000,0.71410000)
 rgb(0.31764706)=(0.41820000,0.59050000,0.71870000)
 rgb(0.32156863)=(0.42830000,0.59800000,0.72330000)
 rgb(0.32549020)=(0.43850000,0.60550000,0.72780000)
 rgb(0.32941176)=(0.44890000,0.61280000,0.73220000)
 rgb(0.33333333)=(0.45930000,0.62010000,0.73640000)
 rgb(0.33725490)=(0.46980000,0.62730000,0.74060000)
 rgb(0.34117647)=(0.48040000,0.63450000,0.74460000)
 rgb(0.34509804)=(0.49100000,0.64150000,0.74850000)
 rgb(0.34901961)=(0.50170000,0.64840000,0.75230000)
 rgb(0.35294118)=(0.51250000,0.65520000,0.75590000)
 rgb(0.35686275)=(0.52330000,0.66200000,0.75930000)
 rgb(0.36078431)=(0.53410000,0.66850000,0.76260000)
 rgb(0.36470588)=(0.54490000,0.67490000,0.76560000)
 rgb(0.36862745)=(0.55580000,0.68120000,0.76850000)
 rgb(0.37254902)=(0.56660000,0.68740000,0.77120000)
 rgb(0.37647059)=(0.57750000,0.69330000,0.77370000)
 rgb(0.38039216)=(0.58830000,0.69910000,0.77600000)
 rgb(0.38431373)=(0.59900000,0.70470000,0.77800000)
 rgb(0.38823529)=(0.60980000,0.71010000,0.77980000)
 rgb(0.39215686)=(0.62040000,0.71530000,0.78140000)
 rgb(0.39607843)=(0.63100000,0.72030000,0.78260000)
 rgb(0.40000000)=(0.64140000,0.72500000,0.78360000)
 rgb(0.40392157)=(0.65180000,0.72950000,0.78440000)
 rgb(0.40784314)=(0.66200000,0.73380000,0.78480000)
 rgb(0.41176471)=(0.67210000,0.73780000,0.78500000)
 rgb(0.41568627)=(0.68200000,0.74150000,0.78480000)
 rgb(0.41960784)=(0.69170000,0.74490000,0.78430000)
 rgb(0.42352941)=(0.70130000,0.74810000,0.78360000)
 rgb(0.42745098)=(0.71060000,0.75100000,0.78250000)
 rgb(0.43137255)=(0.71980000,0.75360000,0.78100000)
 rgb(0.43529412)=(0.72870000,0.75580000,0.77930000)
 rgb(0.43921569)=(0.73730000,0.75780000,0.77720000)
 rgb(0.44313725)=(0.74570000,0.75940000,0.77480000)
 rgb(0.44705882)=(0.75380000,0.76080000,0.77200000)
 rgb(0.45098039)=(0.76160000,0.76180000,0.76900000)
 rgb(0.45490196)=(0.76910000,0.76250000,0.76560000)
 rgb(0.45882353)=(0.77640000,0.76280000,0.76190000)
 rgb(0.46274510)=(0.78320000,0.76290000,0.75780000)
 rgb(0.46666667)=(0.78980000,0.76260000,0.75350000)
 rgb(0.47058824)=(0.79610000,0.76200000,0.74880000)
 rgb(0.47450980)=(0.80200000,0.76110000,0.74390000)
 rgb(0.47843137)=(0.80760000,0.75980000,0.73870000)
 rgb(0.48235294)=(0.81280000,0.75830000,0.73320000)
 rgb(0.48627451)=(0.81770000,0.75650000,0.72740000)
 rgb(0.49019608)=(0.82220000,0.75430000,0.72130000)
 rgb(0.49411765)=(0.82640000,0.75190000,0.71510000)
 rgb(0.49803922)=(0.83030000,0.74920000,0.70860000)
 rgb(0.50196078)=(0.83390000,0.74630000,0.70180000)
 rgb(0.50588235)=(0.83710000,0.74310000,0.69490000)
 rgb(0.50980392)=(0.84000000,0.73960000,0.68770000)
 rgb(0.51372549)=(0.84250000,0.73590000,0.68040000)
 rgb(0.51764706)=(0.84480000,0.73200000,0.67280000)
 rgb(0.52156863)=(0.84670000,0.72780000,0.66520000)
 rgb(0.52549020)=(0.84840000,0.72340000,0.65730000)
 rgb(0.52941176)=(0.84970000,0.71890000,0.64930000)
 rgb(0.53333333)=(0.85080000,0.71410000,0.64120000)
 rgb(0.53725490)=(0.85160000,0.70920000,0.63300000)
 rgb(0.54117647)=(0.85220000,0.70410000,0.62460000)
 rgb(0.54509804)=(0.85250000,0.69880000,0.61620000)
 rgb(0.54901961)=(0.85260000,0.69340000,0.60760000)
 rgb(0.55294118)=(0.85240000,0.68780000,0.59900000)
 rgb(0.55686275)=(0.85200000,0.68210000,0.59030000)
 rgb(0.56078431)=(0.85140000,0.67620000,0.58150000)
 rgb(0.56470588)=(0.85050000,0.67030000,0.57270000)
 rgb(0.56862745)=(0.84950000,0.66420000,0.56380000)
 rgb(0.57254902)=(0.84830000,0.65800000,0.55490000)
 rgb(0.57647059)=(0.84690000,0.65170000,0.54590000)
 rgb(0.58039216)=(0.84530000,0.64530000,0.53690000)
 rgb(0.58431373)=(0.84350000,0.63880000,0.52780000)
 rgb(0.58823529)=(0.84160000,0.63220000,0.51880000)
 rgb(0.59215686)=(0.83950000,0.62550000,0.50970000)
 rgb(0.59607843)=(0.83720000,0.61880000,0.50050000)
 rgb(0.60000000)=(0.83480000,0.61190000,0.49140000)
 rgb(0.60392157)=(0.83230000,0.60500000,0.48230000)
 rgb(0.60784314)=(0.82960000,0.59800000,0.47310000)
 rgb(0.61176471)=(0.82670000,0.59090000,0.46400000)
 rgb(0.61568627)=(0.82370000,0.58380000,0.45480000)
 rgb(0.61960784)=(0.82060000,0.57650000,0.44570000)
 rgb(0.62352941)=(0.81730000,0.56920000,0.43660000)
 rgb(0.62745098)=(0.81390000,0.56190000,0.42750000)
 rgb(0.63137255)=(0.81040000,0.55450000,0.41830000)
 rgb(0.63529412)=(0.80670000,0.54690000,0.40930000)
 rgb(0.63921569)=(0.80290000,0.53940000,0.40020000)
 rgb(0.64313725)=(0.79890000,0.53170000,0.39110000)
 rgb(0.64705882)=(0.79480000,0.52400000,0.38210000)
 rgb(0.65098039)=(0.79060000,0.51620000,0.37310000)
 rgb(0.65490196)=(0.78620000,0.50840000,0.36420000)
 rgb(0.65882353)=(0.78170000,0.50040000,0.35530000)
 rgb(0.66274510)=(0.77710000,0.49240000,0.34640000)
 rgb(0.66666667)=(0.77230000,0.48440000,0.33760000)
 rgb(0.67058824)=(0.76730000,0.47620000,0.32880000)
 rgb(0.67450980)=(0.76220000,0.46800000,0.32010000)
 rgb(0.67843137)=(0.75700000,0.45970000,0.31140000)
 rgb(0.68235294)=(0.75160000,0.45140000,0.30280000)
 rgb(0.68627451)=(0.74600000,0.44300000,0.29430000)
 rgb(0.69019608)=(0.74030000,0.43450000,0.28590000)
 rgb(0.69411765)=(0.73450000,0.42600000,0.27760000)
 rgb(0.69803922)=(0.72850000,0.41740000,0.26940000)
 rgb(0.70196078)=(0.72230000,0.40880000,0.26130000)
 rgb(0.70588235)=(0.71600000,0.40010000,0.25340000)
 rgb(0.70980392)=(0.70960000,0.39140000,0.24550000)
 rgb(0.71372549)=(0.70300000,0.38260000,0.23780000)
 rgb(0.71764706)=(0.69630000,0.37380000,0.23030000)
 rgb(0.72156863)=(0.68940000,0.36500000,0.22290000)
 rgb(0.72549020)=(0.68250000,0.35620000,0.21570000)
 rgb(0.72941176)=(0.67540000,0.34740000,0.20870000)
 rgb(0.73333333)=(0.66820000,0.33850000,0.20190000)
 rgb(0.73725490)=(0.66090000,0.32970000,0.19530000)
 rgb(0.74117647)=(0.65350000,0.32090000,0.18900000)
 rgb(0.74509804)=(0.64600000,0.31220000,0.18290000)
 rgb(0.74901961)=(0.63850000,0.30350000,0.17700000)
 rgb(0.75294118)=(0.63090000,0.29490000,0.17140000)
 rgb(0.75686275)=(0.62330000,0.28630000,0.16600000)
 rgb(0.76078431)=(0.61560000,0.27790000,0.16100000)
 rgb(0.76470588)=(0.60790000,0.26950000,0.15620000)
 rgb(0.76862745)=(0.60020000,0.26120000,0.15160000)
 rgb(0.77254902)=(0.59250000,0.25310000,0.14750000)
 rgb(0.77647059)=(0.58490000,0.24500000,0.14350000)
 rgb(0.78039216)=(0.57720000,0.23720000,0.13990000)
 rgb(0.78431373)=(0.56960000,0.22950000,0.13660000)
 rgb(0.78823529)=(0.56210000,0.22190000,0.13360000)
 rgb(0.79215686)=(0.55460000,0.21450000,0.13080000)
 rgb(0.79607843)=(0.54720000,0.20730000,0.12840000)
 rgb(0.80000000)=(0.53990000,0.20020000,0.12620000)
 rgb(0.80392157)=(0.53270000,0.19340000,0.12430000)
 rgb(0.80784314)=(0.52560000,0.18680000,0.12270000)
 rgb(0.81176471)=(0.51860000,0.18030000,0.12140000)
 rgb(0.81568627)=(0.51170000,0.17410000,0.12030000)
 rgb(0.81960784)=(0.50500000,0.16810000,0.11950000)
 rgb(0.82352941)=(0.49840000,0.16230000,0.11890000)
 rgb(0.82745098)=(0.49190000,0.15670000,0.11850000)
 rgb(0.83137255)=(0.48560000,0.15130000,0.11840000)
 rgb(0.83529412)=(0.47940000,0.14620000,0.11840000)
 rgb(0.83921569)=(0.47330000,0.14130000,0.11870000)
 rgb(0.84313725)=(0.46740000,0.13650000,0.11910000)
 rgb(0.84705882)=(0.46160000,0.13200000,0.11970000)
 rgb(0.85098039)=(0.45600000,0.12770000,0.12050000)
 rgb(0.85490196)=(0.45050000,0.12360000,0.12150000)
 rgb(0.85882353)=(0.44510000,0.11980000,0.12260000)
 rgb(0.86274510)=(0.43990000,0.11620000,0.12390000)
 rgb(0.86666667)=(0.43480000,0.11280000,0.12540000)
 rgb(0.87058824)=(0.42980000,0.10960000,0.12690000)
 rgb(0.87450980)=(0.42500000,0.10650000,0.12860000)
 rgb(0.87843137)=(0.42030000,0.10370000,0.13050000)
 rgb(0.88235294)=(0.41570000,0.10110000,0.13240000)
 rgb(0.88627451)=(0.41120000,0.09870000,0.13450000)
 rgb(0.89019608)=(0.40680000,0.09640000,0.13660000)
 rgb(0.89411765)=(0.40260000,0.09440000,0.13890000)
 rgb(0.89803922)=(0.39840000,0.09250000,0.14130000)
 rgb(0.90196078)=(0.39430000,0.09090000,0.14380000)
 rgb(0.90588235)=(0.39040000,0.08940000,0.14640000)
 rgb(0.90980392)=(0.38650000,0.08810000,0.14900000)
 rgb(0.91372549)=(0.38270000,0.08700000,0.15180000)
 rgb(0.91764706)=(0.37900000,0.08590000,0.15470000)
 rgb(0.92156863)=(0.37530000,0.08510000,0.15760000)
 rgb(0.92549020)=(0.37170000,0.08450000,0.16060000)
 rgb(0.92941176)=(0.36820000,0.08410000,0.16380000)
 rgb(0.93333333)=(0.36480000,0.08370000,0.16700000)
 rgb(0.93725490)=(0.36140000,0.08350000,0.17030000)
 rgb(0.94117647)=(0.35810000,0.08350000,0.17360000)
 rgb(0.94509804)=(0.35480000,0.08360000,0.17710000)
 rgb(0.94901961)=(0.35160000,0.08390000,0.18060000)
 rgb(0.95294118)=(0.34840000,0.08430000,0.18420000)
 rgb(0.95686275)=(0.34530000,0.08480000,0.18790000)
 rgb(0.96078431)=(0.34220000,0.08540000,0.19170000)
 rgb(0.96470588)=(0.33910000,0.08620000,0.19550000)
 rgb(0.96862745)=(0.33610000,0.08720000,0.19940000)
 rgb(0.97254902)=(0.33310000,0.08820000,0.20350000)
 rgb(0.97647059)=(0.33010000,0.08940000,0.20750000)
 rgb(0.98039216)=(0.32720000,0.09070000,0.21170000)
 rgb(0.98431373)=(0.32420000,0.09210000,0.21600000)
 rgb(0.98823529)=(0.32130000,0.09360000,0.22030000)
 rgb(0.99215686)=(0.31840000,0.09530000,0.22470000)
 rgb(0.99607843)=(0.31550000,0.09700000,0.22920000)
 rgb(1.00000000)=(0.31260000,0.09890000,0.23380000)},
 }

 \pgfplotsset{
 colormap={plots2}{rgb(0.00000000)=(0.00130000,0.06980000,0.37950000)
 rgb(0.00392157)=(0.00240000,0.07650000,0.38350000)
 rgb(0.00784314)=(0.00330000,0.08310000,0.38750000)
 rgb(0.01176471)=(0.00410000,0.08960000,0.39150000)
 rgb(0.01568627)=(0.00490000,0.09590000,0.39550000)
 rgb(0.01960784)=(0.00560000,0.10230000,0.39940000)
 rgb(0.02352941)=(0.00620000,0.10850000,0.40340000)
 rgb(0.02745098)=(0.00670000,0.11470000,0.40730000)
 rgb(0.03137255)=(0.00710000,0.12080000,0.41130000)
 rgb(0.03529412)=(0.00750000,0.12700000,0.41520000)
 rgb(0.03921569)=(0.00780000,0.13310000,0.41920000)
 rgb(0.04313725)=(0.00810000,0.13910000,0.42310000)
 rgb(0.04705882)=(0.00840000,0.14520000,0.42700000)
 rgb(0.05098039)=(0.00860000,0.15110000,0.43090000)
 rgb(0.05490196)=(0.00880000,0.15710000,0.43480000)
 rgb(0.05882353)=(0.00890000,0.16320000,0.43870000)
 rgb(0.06274510)=(0.00910000,0.16910000,0.44260000)
 rgb(0.06666667)=(0.00920000,0.17510000,0.44650000)
 rgb(0.07058824)=(0.00930000,0.18110000,0.45030000)
 rgb(0.07450980)=(0.00940000,0.18710000,0.45420000)
 rgb(0.07843137)=(0.00940000,0.19300000,0.45810000)
 rgb(0.08235294)=(0.00950000,0.19900000,0.46200000)
 rgb(0.08627451)=(0.00960000,0.20500000,0.46580000)
 rgb(0.09019608)=(0.00960000,0.21100000,0.46970000)
 rgb(0.09411765)=(0.00970000,0.21700000,0.47360000)
 rgb(0.09803922)=(0.00970000,0.22310000,0.47750000)
 rgb(0.10196078)=(0.00980000,0.22910000,0.48140000)
 rgb(0.10588235)=(0.00990000,0.23520000,0.48520000)
 rgb(0.10980392)=(0.01000000,0.24130000,0.48920000)
 rgb(0.11372549)=(0.01010000,0.24740000,0.49310000)
 rgb(0.11764706)=(0.01030000,0.25350000,0.49700000)
 rgb(0.12156863)=(0.01050000,0.25970000,0.50100000)
 rgb(0.12549020)=(0.01080000,0.26590000,0.50490000)
 rgb(0.12941176)=(0.01120000,0.27200000,0.50890000)
 rgb(0.13333333)=(0.01170000,0.27830000,0.51290000)
 rgb(0.13725490)=(0.01230000,0.28460000,0.51700000)
 rgb(0.14117647)=(0.01290000,0.29090000,0.52100000)
 rgb(0.14509804)=(0.01380000,0.29720000,0.52510000)
 rgb(0.14901961)=(0.01480000,0.30360000,0.52920000)
 rgb(0.15294118)=(0.01610000,0.31000000,0.53330000)
 rgb(0.15686275)=(0.01770000,0.31650000,0.53750000)
 rgb(0.16078431)=(0.01960000,0.32300000,0.54170000)
 rgb(0.16470588)=(0.02190000,0.32960000,0.54590000)
 rgb(0.16862745)=(0.02470000,0.33610000,0.55020000)
 rgb(0.17254902)=(0.02800000,0.34280000,0.55450000)
 rgb(0.17647059)=(0.03200000,0.34950000,0.55890000)
 rgb(0.18039216)=(0.03680000,0.35630000,0.56330000)
 rgb(0.18431373)=(0.04220000,0.36320000,0.56780000)
 rgb(0.18823529)=(0.04800000,0.37010000,0.57230000)
 rgb(0.19215686)=(0.05430000,0.37710000,0.57690000)
 rgb(0.19607843)=(0.06100000,0.38410000,0.58160000)
 rgb(0.20000000)=(0.06810000,0.39130000,0.58630000)
 rgb(0.20392157)=(0.07550000,0.39850000,0.59100000)
 rgb(0.20784314)=(0.08320000,0.40570000,0.59590000)
 rgb(0.21176471)=(0.09140000,0.41310000,0.60080000)
 rgb(0.21568627)=(0.09980000,0.42050000,0.60570000)
 rgb(0.21960784)=(0.10860000,0.42800000,0.61070000)
 rgb(0.22352941)=(0.11770000,0.43560000,0.61580000)
 rgb(0.22745098)=(0.12700000,0.44320000,0.62090000)
 rgb(0.23137255)=(0.13670000,0.45090000,0.62610000)
 rgb(0.23529412)=(0.14660000,0.45860000,0.63130000)
 rgb(0.23921569)=(0.15680000,0.46650000,0.63660000)
 rgb(0.24313725)=(0.16720000,0.47430000,0.64190000)
 rgb(0.24705882)=(0.17780000,0.48220000,0.64720000)
 rgb(0.25098039)=(0.18860000,0.49020000,0.65260000)
 rgb(0.25490196)=(0.19960000,0.49820000,0.65800000)
 rgb(0.25882353)=(0.21080000,0.50620000,0.66350000)
 rgb(0.26274510)=(0.22210000,0.51430000,0.66890000)
 rgb(0.26666667)=(0.23360000,0.52230000,0.67440000)
 rgb(0.27058824)=(0.24520000,0.53040000,0.67990000)
 rgb(0.27450980)=(0.25700000,0.53850000,0.68540000)
 rgb(0.27843137)=(0.26890000,0.54660000,0.69090000)
 rgb(0.28235294)=(0.28080000,0.55470000,0.69640000)
 rgb(0.28627451)=(0.29290000,0.56280000,0.70190000)
 rgb(0.29019608)=(0.30500000,0.57090000,0.70740000)
 rgb(0.29411765)=(0.31720000,0.57900000,0.71300000)
 rgb(0.29803922)=(0.32940000,0.58710000,0.71840000)
 rgb(0.30196078)=(0.34170000,0.59510000,0.72390000)
 rgb(0.30588235)=(0.35410000,0.60320000,0.72940000)
 rgb(0.30980392)=(0.36650000,0.61120000,0.73490000)
 rgb(0.31372549)=(0.37890000,0.61920000,0.74030000)
 rgb(0.31764706)=(0.39130000,0.62720000,0.74580000)
 rgb(0.32156863)=(0.40380000,0.63510000,0.75120000)
 rgb(0.32549020)=(0.41620000,0.64300000,0.75660000)
 rgb(0.32941176)=(0.42870000,0.65100000,0.76200000)
 rgb(0.33333333)=(0.44120000,0.65880000,0.76730000)
 rgb(0.33725490)=(0.45370000,0.66670000,0.77270000)
 rgb(0.34117647)=(0.46620000,0.67450000,0.77800000)
 rgb(0.34509804)=(0.47870000,0.68230000,0.78340000)
 rgb(0.34901961)=(0.49120000,0.69010000,0.78870000)
 rgb(0.35294118)=(0.50370000,0.69790000,0.79400000)
 rgb(0.35686275)=(0.51620000,0.70570000,0.79930000)
 rgb(0.36078431)=(0.52870000,0.71340000,0.80450000)
 rgb(0.36470588)=(0.54110000,0.72110000,0.80980000)
 rgb(0.36862745)=(0.55360000,0.72880000,0.81500000)
 rgb(0.37254902)=(0.56610000,0.73640000,0.82020000)
 rgb(0.37647059)=(0.57860000,0.74410000,0.82540000)
 rgb(0.38039216)=(0.59100000,0.75170000,0.83060000)
 rgb(0.38431373)=(0.60350000,0.75930000,0.83580000)
 rgb(0.38823529)=(0.61590000,0.76690000,0.84090000)
 rgb(0.39215686)=(0.62840000,0.77450000,0.84610000)
 rgb(0.39607843)=(0.64080000,0.78200000,0.85110000)
 rgb(0.40000000)=(0.65320000,0.78950000,0.85620000)
 rgb(0.40392157)=(0.66560000,0.79690000,0.86120000)
 rgb(0.40784314)=(0.67810000,0.80440000,0.86620000)
 rgb(0.41176471)=(0.69050000,0.81170000,0.87110000)
 rgb(0.41568627)=(0.70290000,0.81900000,0.87590000)
 rgb(0.41960784)=(0.71530000,0.82630000,0.88060000)
 rgb(0.42352941)=(0.72760000,0.83340000,0.88510000)
 rgb(0.42745098)=(0.74000000,0.84050000,0.88960000)
 rgb(0.43137255)=(0.75240000,0.84740000,0.89380000)
 rgb(0.43529412)=(0.76470000,0.85410000,0.89780000)
 rgb(0.43921569)=(0.77690000,0.86070000,0.90160000)
 rgb(0.44313725)=(0.78910000,0.86700000,0.90500000)
 rgb(0.44705882)=(0.80120000,0.87300000,0.90800000)
 rgb(0.45098039)=(0.81310000,0.87870000,0.91070000)
 rgb(0.45490196)=(0.82490000,0.88410000,0.91280000)
 rgb(0.45882353)=(0.83640000,0.88890000,0.91430000)
 rgb(0.46274510)=(0.84760000,0.89330000,0.91520000)
 rgb(0.46666667)=(0.85850000,0.89710000,0.91540000)
 rgb(0.47058824)=(0.86890000,0.90020000,0.91480000)
 rgb(0.47450980)=(0.87870000,0.90260000,0.91340000)
 rgb(0.47843137)=(0.88800000,0.90430000,0.91120000)
 rgb(0.48235294)=(0.89650000,0.90520000,0.90800000)
 rgb(0.48627451)=(0.90420000,0.90520000,0.90400000)
 rgb(0.49019608)=(0.91120000,0.90440000,0.89910000)
 rgb(0.49411765)=(0.91720000,0.90280000,0.89340000)
 rgb(0.49803922)=(0.92230000,0.90040000,0.88690000)
 rgb(0.50196078)=(0.92650000,0.89720000,0.87970000)
 rgb(0.50588235)=(0.92980000,0.89330000,0.87180000)
 rgb(0.50980392)=(0.93220000,0.88870000,0.86340000)
 rgb(0.51372549)=(0.93390000,0.88360000,0.85450000)
 rgb(0.51764706)=(0.93480000,0.87790000,0.84520000)
 rgb(0.52156863)=(0.93500000,0.87180000,0.83550000)
 rgb(0.52549020)=(0.93460000,0.86530000,0.82560000)
 rgb(0.52941176)=(0.93380000,0.85850000,0.81540000)
 rgb(0.53333333)=(0.93240000,0.85150000,0.80510000)
 rgb(0.53725490)=(0.93070000,0.84420000,0.79470000)
 rgb(0.54117647)=(0.92860000,0.83680000,0.78420000)
 rgb(0.54509804)=(0.92630000,0.82920000,0.77360000)
 rgb(0.54901961)=(0.92380000,0.82150000,0.76300000)
 rgb(0.55294118)=(0.92100000,0.81380000,0.75230000)
 rgb(0.55686275)=(0.91810000,0.80600000,0.74170000)
 rgb(0.56078431)=(0.91520000,0.79820000,0.73100000)
 rgb(0.56470588)=(0.91210000,0.79030000,0.72040000)
 rgb(0.56862745)=(0.90890000,0.78240000,0.70980000)
 rgb(0.57254902)=(0.90570000,0.77450000,0.69920000)
 rgb(0.57647059)=(0.90250000,0.76670000,0.68860000)
 rgb(0.58039216)=(0.89920000,0.75880000,0.67810000)
 rgb(0.58431373)=(0.89600000,0.75100000,0.66760000)
 rgb(0.58823529)=(0.89270000,0.74310000,0.65710000)
 rgb(0.59215686)=(0.88940000,0.73530000,0.64670000)
 rgb(0.59607843)=(0.88610000,0.72760000,0.63630000)
 rgb(0.60000000)=(0.88280000,0.71980000,0.62590000)
 rgb(0.60392157)=(0.87960000,0.71210000,0.61560000)
 rgb(0.60784314)=(0.87630000,0.70440000,0.60540000)
 rgb(0.61176471)=(0.87300000,0.69680000,0.59510000)
 rgb(0.61568627)=(0.86980000,0.68910000,0.58500000)
 rgb(0.61960784)=(0.86660000,0.68150000,0.57480000)
 rgb(0.62352941)=(0.86330000,0.67400000,0.56470000)
 rgb(0.62745098)=(0.86010000,0.66650000,0.55470000)
 rgb(0.63137255)=(0.85690000,0.65900000,0.54470000)
 rgb(0.63529412)=(0.85370000,0.65150000,0.53480000)
 rgb(0.63921569)=(0.85060000,0.64410000,0.52480000)
 rgb(0.64313725)=(0.84740000,0.63670000,0.51500000)
 rgb(0.64705882)=(0.84430000,0.62930000,0.50510000)
 rgb(0.65098039)=(0.84110000,0.62200000,0.49540000)
 rgb(0.65490196)=(0.83800000,0.61470000,0.48560000)
 rgb(0.65882353)=(0.83490000,0.60740000,0.47590000)
 rgb(0.66274510)=(0.83180000,0.60010000,0.46630000)
 rgb(0.66666667)=(0.82870000,0.59290000,0.45670000)
 rgb(0.67058824)=(0.82560000,0.58580000,0.44710000)
 rgb(0.67450980)=(0.82260000,0.57860000,0.43760000)
 rgb(0.67843137)=(0.81950000,0.57150000,0.42810000)
 rgb(0.68235294)=(0.81650000,0.56440000,0.41870000)
 rgb(0.68627451)=(0.81350000,0.55730000,0.40930000)
 rgb(0.69019608)=(0.81040000,0.55030000,0.39990000)
 rgb(0.69411765)=(0.80740000,0.54330000,0.39060000)
 rgb(0.69803922)=(0.80440000,0.53630000,0.38130000)
 rgb(0.70196078)=(0.80150000,0.52930000,0.37200000)
 rgb(0.70588235)=(0.79850000,0.52240000,0.36280000)
 rgb(0.70980392)=(0.79550000,0.51550000,0.35370000)
 rgb(0.71372549)=(0.79250000,0.50860000,0.34450000)
 rgb(0.71764706)=(0.78960000,0.50170000,0.33540000)
 rgb(0.72156863)=(0.78660000,0.49480000,0.32630000)
 rgb(0.72549020)=(0.78370000,0.48800000,0.31730000)
 rgb(0.72941176)=(0.78070000,0.48110000,0.30830000)
 rgb(0.73333333)=(0.77770000,0.47430000,0.29930000)
 rgb(0.73725490)=(0.77480000,0.46750000,0.29040000)
 rgb(0.74117647)=(0.77180000,0.46060000,0.28140000)
 rgb(0.74509804)=(0.76880000,0.45380000,0.27250000)
 rgb(0.74901961)=(0.76580000,0.44690000,0.26360000)
 rgb(0.75294118)=(0.76270000,0.44010000,0.25480000)
 rgb(0.75686275)=(0.75960000,0.43310000,0.24590000)
 rgb(0.76078431)=(0.75650000,0.42620000,0.23700000)
 rgb(0.76470588)=(0.75330000,0.41920000,0.22820000)
 rgb(0.76862745)=(0.75010000,0.41220000,0.21930000)
 rgb(0.77254902)=(0.74670000,0.40500000,0.21050000)
 rgb(0.77647059)=(0.74320000,0.39780000,0.20160000)
 rgb(0.78039216)=(0.73970000,0.39050000,0.19270000)
 rgb(0.78431373)=(0.73590000,0.38310000,0.18390000)
 rgb(0.78823529)=(0.73200000,0.37550000,0.17500000)
 rgb(0.79215686)=(0.72790000,0.36770000,0.16600000)
 rgb(0.79607843)=(0.72350000,0.35990000,0.15710000)
 rgb(0.80000000)=(0.71890000,0.35180000,0.14820000)
 rgb(0.80392157)=(0.71400000,0.34350000,0.13930000)
 rgb(0.80784314)=(0.70880000,0.33500000,0.13050000)
 rgb(0.81176471)=(0.70330000,0.32640000,0.12150000)
 rgb(0.81568627)=(0.69740000,0.31750000,0.11280000)
 rgb(0.81960784)=(0.69120000,0.30850000,0.10410000)
 rgb(0.82352941)=(0.68470000,0.29930000,0.09560000)
 rgb(0.82745098)=(0.67770000,0.28990000,0.08740000)
 rgb(0.83137255)=(0.67050000,0.28050000,0.07920000)
 rgb(0.83529412)=(0.66290000,0.27100000,0.07150000)
 rgb(0.83921569)=(0.65500000,0.26150000,0.06410000)
 rgb(0.84313725)=(0.64700000,0.25210000,0.05710000)
 rgb(0.84705882)=(0.63870000,0.24270000,0.05060000)
 rgb(0.85098039)=(0.63030000,0.23350000,0.04480000)
 rgb(0.85490196)=(0.62170000,0.22440000,0.03940000)
 rgb(0.85882353)=(0.61310000,0.21570000,0.03480000)
 rgb(0.86274510)=(0.60450000,0.20710000,0.03110000)
 rgb(0.86666667)=(0.59590000,0.19870000,0.02820000)
 rgb(0.87058824)=(0.58740000,0.19070000,0.02600000)
 rgb(0.87450980)=(0.57890000,0.18290000,0.02440000)
 rgb(0.87843137)=(0.57050000,0.17540000,0.02330000)
 rgb(0.88235294)=(0.56230000,0.16820000,0.02250000)
 rgb(0.88627451)=(0.55410000,0.16120000,0.02210000)
 rgb(0.89019608)=(0.54600000,0.15440000,0.02190000)
 rgb(0.89411765)=(0.53800000,0.14790000,0.02170000)
 rgb(0.89803922)=(0.53020000,0.14150000,0.02170000)
 rgb(0.90196078)=(0.52240000,0.13530000,0.02180000)
 rgb(0.90588235)=(0.51480000,0.12920000,0.02200000)
 rgb(0.90980392)=(0.50720000,0.12330000,0.02220000)
 rgb(0.91372549)=(0.49970000,0.11750000,0.02250000)
 rgb(0.91764706)=(0.49230000,0.11180000,0.02280000)
 rgb(0.92156863)=(0.48500000,0.10620000,0.02310000)
 rgb(0.92549020)=(0.47780000,0.10060000,0.02350000)
 rgb(0.92941176)=(0.47060000,0.09520000,0.02390000)
 rgb(0.93333333)=(0.46350000,0.08970000,0.02430000)
 rgb(0.93725490)=(0.45650000,0.08430000,0.02480000)
 rgb(0.94117647)=(0.44950000,0.07870000,0.02520000)
 rgb(0.94509804)=(0.44260000,0.07340000,0.02560000)
 rgb(0.94901961)=(0.43570000,0.06790000,0.02610000)
 rgb(0.95294118)=(0.42890000,0.06240000,0.02650000)
 rgb(0.95686275)=(0.42210000,0.05680000,0.02700000)
 rgb(0.96078431)=(0.41540000,0.05110000,0.02740000)
 rgb(0.96470588)=(0.40880000,0.04540000,0.02780000)
 rgb(0.96862745)=(0.40210000,0.03940000,0.02820000)
 rgb(0.97254902)=(0.39560000,0.03340000,0.02860000)
 rgb(0.97647059)=(0.38900000,0.02780000,0.02890000)
 rgb(0.98039216)=(0.38250000,0.02260000,0.02930000)
 rgb(0.98431373)=(0.37600000,0.01760000,0.02960000)
 rgb(0.98823529)=(0.36960000,0.01290000,0.02990000)
 rgb(0.99215686)=(0.36320000,0.00820000,0.03010000)
 rgb(0.99607843)=(0.35680000,0.00400000,0.03030000)
 rgb(1.00000000)=(0.35040000,0.00010000,0.03050000)},
 }
 \pgfplotsset{
 colormap={plots2}{rgb(0.00000000)=(0.00130000,0.06980000,0.37950000)
 rgb(0.00392157)=(0.00240000,0.07650000,0.38350000)
 rgb(0.00784314)=(0.00330000,0.08310000,0.38750000)
 rgb(0.01176471)=(0.00410000,0.08960000,0.39150000)
 rgb(0.01568627)=(0.00490000,0.09590000,0.39550000)
 rgb(0.01960784)=(0.00560000,0.10230000,0.39940000)
 rgb(0.02352941)=(0.00620000,0.10850000,0.40340000)
 rgb(0.02745098)=(0.00670000,0.11470000,0.40730000)
 rgb(0.03137255)=(0.00710000,0.12080000,0.41130000)
 rgb(0.03529412)=(0.00750000,0.12700000,0.41520000)
 rgb(0.03921569)=(0.00780000,0.13310000,0.41920000)
 rgb(0.04313725)=(0.00810000,0.13910000,0.42310000)
 rgb(0.04705882)=(0.00840000,0.14520000,0.42700000)
 rgb(0.05098039)=(0.00860000,0.15110000,0.43090000)
 rgb(0.05490196)=(0.00880000,0.15710000,0.43480000)
 rgb(0.05882353)=(0.00890000,0.16320000,0.43870000)
 rgb(0.06274510)=(0.00910000,0.16910000,0.44260000)
 rgb(0.06666667)=(0.00920000,0.17510000,0.44650000)
 rgb(0.07058824)=(0.00930000,0.18110000,0.45030000)
 rgb(0.07450980)=(0.00940000,0.18710000,0.45420000)
 rgb(0.07843137)=(0.00940000,0.19300000,0.45810000)
 rgb(0.08235294)=(0.00950000,0.19900000,0.46200000)
 rgb(0.08627451)=(0.00960000,0.20500000,0.46580000)
 rgb(0.09019608)=(0.00960000,0.21100000,0.46970000)
 rgb(0.09411765)=(0.00970000,0.21700000,0.47360000)
 rgb(0.09803922)=(0.00970000,0.22310000,0.47750000)
 rgb(0.10196078)=(0.00980000,0.22910000,0.48140000)
 rgb(0.10588235)=(0.00990000,0.23520000,0.48520000)
 rgb(0.10980392)=(0.01000000,0.24130000,0.48920000)
 rgb(0.11372549)=(0.01010000,0.24740000,0.49310000)
 rgb(0.11764706)=(0.01030000,0.25350000,0.49700000)
 rgb(0.12156863)=(0.01050000,0.25970000,0.50100000)
 rgb(0.12549020)=(0.01080000,0.26590000,0.50490000)
 rgb(0.12941176)=(0.01120000,0.27200000,0.50890000)
 rgb(0.13333333)=(0.01170000,0.27830000,0.51290000)
 rgb(0.13725490)=(0.01230000,0.28460000,0.51700000)
 rgb(0.14117647)=(0.01290000,0.29090000,0.52100000)
 rgb(0.14509804)=(0.01380000,0.29720000,0.52510000)
 rgb(0.14901961)=(0.01480000,0.30360000,0.52920000)
 rgb(0.15294118)=(0.01610000,0.31000000,0.53330000)
 rgb(0.15686275)=(0.01770000,0.31650000,0.53750000)
 rgb(0.16078431)=(0.01960000,0.32300000,0.54170000)
 rgb(0.16470588)=(0.02190000,0.32960000,0.54590000)
 rgb(0.16862745)=(0.02470000,0.33610000,0.55020000)
 rgb(0.17254902)=(0.02800000,0.34280000,0.55450000)
 rgb(0.17647059)=(0.03200000,0.34950000,0.55890000)
 rgb(0.18039216)=(0.03680000,0.35630000,0.56330000)
 rgb(0.18431373)=(0.04220000,0.36320000,0.56780000)
 rgb(0.18823529)=(0.04800000,0.37010000,0.57230000)
 rgb(0.19215686)=(0.05430000,0.37710000,0.57690000)
 rgb(0.19607843)=(0.06100000,0.38410000,0.58160000)
 rgb(0.20000000)=(0.06810000,0.39130000,0.58630000)
 rgb(0.20392157)=(0.07550000,0.39850000,0.59100000)
 rgb(0.20784314)=(0.08320000,0.40570000,0.59590000)
 rgb(0.21176471)=(0.09140000,0.41310000,0.60080000)
 rgb(0.21568627)=(0.09980000,0.42050000,0.60570000)
 rgb(0.21960784)=(0.10860000,0.42800000,0.61070000)
 rgb(0.22352941)=(0.11770000,0.43560000,0.61580000)
 rgb(0.22745098)=(0.12700000,0.44320000,0.62090000)
 rgb(0.23137255)=(0.13670000,0.45090000,0.62610000)
 rgb(0.23529412)=(0.14660000,0.45860000,0.63130000)
 rgb(0.23921569)=(0.15680000,0.46650000,0.63660000)
 rgb(0.24313725)=(0.16720000,0.47430000,0.64190000)
 rgb(0.24705882)=(0.17780000,0.48220000,0.64720000)
 rgb(0.25098039)=(0.18860000,0.49020000,0.65260000)
 rgb(0.25490196)=(0.19960000,0.49820000,0.65800000)
 rgb(0.25882353)=(0.21080000,0.50620000,0.66350000)
 rgb(0.26274510)=(0.22210000,0.51430000,0.66890000)
 rgb(0.26666667)=(0.23360000,0.52230000,0.67440000)
 rgb(0.27058824)=(0.24520000,0.53040000,0.67990000)
 rgb(0.27450980)=(0.25700000,0.53850000,0.68540000)
 rgb(0.27843137)=(0.26890000,0.54660000,0.69090000)
 rgb(0.28235294)=(0.28080000,0.55470000,0.69640000)
 rgb(0.28627451)=(0.29290000,0.56280000,0.70190000)
 rgb(0.29019608)=(0.30500000,0.57090000,0.70740000)
 rgb(0.29411765)=(0.31720000,0.57900000,0.71300000)
 rgb(0.29803922)=(0.32940000,0.58710000,0.71840000)
 rgb(0.30196078)=(0.34170000,0.59510000,0.72390000)
 rgb(0.30588235)=(0.35410000,0.60320000,0.72940000)
 rgb(0.30980392)=(0.36650000,0.61120000,0.73490000)
 rgb(0.31372549)=(0.37890000,0.61920000,0.74030000)
 rgb(0.31764706)=(0.39130000,0.62720000,0.74580000)
 rgb(0.32156863)=(0.40380000,0.63510000,0.75120000)
 rgb(0.32549020)=(0.41620000,0.64300000,0.75660000)
 rgb(0.32941176)=(0.42870000,0.65100000,0.76200000)
 rgb(0.33333333)=(0.44120000,0.65880000,0.76730000)
 rgb(0.33725490)=(0.45370000,0.66670000,0.77270000)
 rgb(0.34117647)=(0.46620000,0.67450000,0.77800000)
 rgb(0.34509804)=(0.47870000,0.68230000,0.78340000)
 rgb(0.34901961)=(0.49120000,0.69010000,0.78870000)
 rgb(0.35294118)=(0.50370000,0.69790000,0.79400000)
 rgb(0.35686275)=(0.51620000,0.70570000,0.79930000)
 rgb(0.36078431)=(0.52870000,0.71340000,0.80450000)
 rgb(0.36470588)=(0.54110000,0.72110000,0.80980000)
 rgb(0.36862745)=(0.55360000,0.72880000,0.81500000)
 rgb(0.37254902)=(0.56610000,0.73640000,0.82020000)
 rgb(0.37647059)=(0.57860000,0.74410000,0.82540000)
 rgb(0.38039216)=(0.59100000,0.75170000,0.83060000)
 rgb(0.38431373)=(0.60350000,0.75930000,0.83580000)
 rgb(0.38823529)=(0.61590000,0.76690000,0.84090000)
 rgb(0.39215686)=(0.62840000,0.77450000,0.84610000)
 rgb(0.39607843)=(0.64080000,0.78200000,0.85110000)
 rgb(0.40000000)=(0.65320000,0.78950000,0.85620000)
 rgb(0.40392157)=(0.66560000,0.79690000,0.86120000)
 rgb(0.40784314)=(0.67810000,0.80440000,0.86620000)
 rgb(0.41176471)=(0.69050000,0.81170000,0.87110000)
 rgb(0.41568627)=(0.70290000,0.81900000,0.87590000)
 rgb(0.41960784)=(0.71530000,0.82630000,0.88060000)
 rgb(0.42352941)=(0.72760000,0.83340000,0.88510000)
 rgb(0.42745098)=(0.74000000,0.84050000,0.88960000)
 rgb(0.43137255)=(0.75240000,0.84740000,0.89380000)
 rgb(0.43529412)=(0.76470000,0.85410000,0.89780000)
 rgb(0.43921569)=(0.77690000,0.86070000,0.90160000)
 rgb(0.44313725)=(0.78910000,0.86700000,0.90500000)
 rgb(0.44705882)=(0.80120000,0.87300000,0.90800000)
 rgb(0.45098039)=(0.81310000,0.87870000,0.91070000)
 rgb(0.45490196)=(0.82490000,0.88410000,0.91280000)
 rgb(0.45882353)=(0.83640000,0.88890000,0.91430000)
 rgb(0.46274510)=(0.84760000,0.89330000,0.91520000)
 rgb(0.46666667)=(0.85850000,0.89710000,0.91540000)
 rgb(0.47058824)=(0.86890000,0.90020000,0.91480000)
 rgb(0.47450980)=(0.87870000,0.90260000,0.91340000)
 rgb(0.47843137)=(0.88800000,0.90430000,0.91120000)
 rgb(0.48235294)=(0.89650000,0.90520000,0.90800000)
 rgb(0.48627451)=(0.90420000,0.90520000,0.90400000)
 rgb(0.49019608)=(0.91120000,0.90440000,0.89910000)
 rgb(0.49411765)=(0.91720000,0.90280000,0.89340000)
 rgb(0.49803922)=(0.92230000,0.90040000,0.88690000)
 rgb(0.50196078)=(0.92650000,0.89720000,0.87970000)
 rgb(0.50588235)=(0.92980000,0.89330000,0.87180000)
 rgb(0.50980392)=(0.93220000,0.88870000,0.86340000)
 rgb(0.51372549)=(0.93390000,0.88360000,0.85450000)
 rgb(0.51764706)=(0.93480000,0.87790000,0.84520000)
 rgb(0.52156863)=(0.93500000,0.87180000,0.83550000)
 rgb(0.52549020)=(0.93460000,0.86530000,0.82560000)
 rgb(0.52941176)=(0.93380000,0.85850000,0.81540000)
 rgb(0.53333333)=(0.93240000,0.85150000,0.80510000)
 rgb(0.53725490)=(0.93070000,0.84420000,0.79470000)
 rgb(0.54117647)=(0.92860000,0.83680000,0.78420000)
 rgb(0.54509804)=(0.92630000,0.82920000,0.77360000)
 rgb(0.54901961)=(0.92380000,0.82150000,0.76300000)
 rgb(0.55294118)=(0.92100000,0.81380000,0.75230000)
 rgb(0.55686275)=(0.91810000,0.80600000,0.74170000)
 rgb(0.56078431)=(0.91520000,0.79820000,0.73100000)
 rgb(0.56470588)=(0.91210000,0.79030000,0.72040000)
 rgb(0.56862745)=(0.90890000,0.78240000,0.70980000)
 rgb(0.57254902)=(0.90570000,0.77450000,0.69920000)
 rgb(0.57647059)=(0.90250000,0.76670000,0.68860000)
 rgb(0.58039216)=(0.89920000,0.75880000,0.67810000)
 rgb(0.58431373)=(0.89600000,0.75100000,0.66760000)
 rgb(0.58823529)=(0.89270000,0.74310000,0.65710000)
 rgb(0.59215686)=(0.88940000,0.73530000,0.64670000)
 rgb(0.59607843)=(0.88610000,0.72760000,0.63630000)
 rgb(0.60000000)=(0.88280000,0.71980000,0.62590000)
 rgb(0.60392157)=(0.87960000,0.71210000,0.61560000)
 rgb(0.60784314)=(0.87630000,0.70440000,0.60540000)
 rgb(0.61176471)=(0.87300000,0.69680000,0.59510000)
 rgb(0.61568627)=(0.86980000,0.68910000,0.58500000)
 rgb(0.61960784)=(0.86660000,0.68150000,0.57480000)
 rgb(0.62352941)=(0.86330000,0.67400000,0.56470000)
 rgb(0.62745098)=(0.86010000,0.66650000,0.55470000)
 rgb(0.63137255)=(0.85690000,0.65900000,0.54470000)
 rgb(0.63529412)=(0.85370000,0.65150000,0.53480000)
 rgb(0.63921569)=(0.85060000,0.64410000,0.52480000)
 rgb(0.64313725)=(0.84740000,0.63670000,0.51500000)
 rgb(0.64705882)=(0.84430000,0.62930000,0.50510000)
 rgb(0.65098039)=(0.84110000,0.62200000,0.49540000)
 rgb(0.65490196)=(0.83800000,0.61470000,0.48560000)
 rgb(0.65882353)=(0.83490000,0.60740000,0.47590000)
 rgb(0.66274510)=(0.83180000,0.60010000,0.46630000)
 rgb(0.66666667)=(0.82870000,0.59290000,0.45670000)
 rgb(0.67058824)=(0.82560000,0.58580000,0.44710000)
 rgb(0.67450980)=(0.82260000,0.57860000,0.43760000)
 rgb(0.67843137)=(0.81950000,0.57150000,0.42810000)
 rgb(0.68235294)=(0.81650000,0.56440000,0.41870000)
 rgb(0.68627451)=(0.81350000,0.55730000,0.40930000)
 rgb(0.69019608)=(0.81040000,0.55030000,0.39990000)
 rgb(0.69411765)=(0.80740000,0.54330000,0.39060000)
 rgb(0.69803922)=(0.80440000,0.53630000,0.38130000)
 rgb(0.70196078)=(0.80150000,0.52930000,0.37200000)
 rgb(0.70588235)=(0.79850000,0.52240000,0.36280000)
 rgb(0.70980392)=(0.79550000,0.51550000,0.35370000)
 rgb(0.71372549)=(0.79250000,0.50860000,0.34450000)
 rgb(0.71764706)=(0.78960000,0.50170000,0.33540000)
 rgb(0.72156863)=(0.78660000,0.49480000,0.32630000)
 rgb(0.72549020)=(0.78370000,0.48800000,0.31730000)
 rgb(0.72941176)=(0.78070000,0.48110000,0.30830000)
 rgb(0.73333333)=(0.77770000,0.47430000,0.29930000)
 rgb(0.73725490)=(0.77480000,0.46750000,0.29040000)
 rgb(0.74117647)=(0.77180000,0.46060000,0.28140000)
 rgb(0.74509804)=(0.76880000,0.45380000,0.27250000)
 rgb(0.74901961)=(0.76580000,0.44690000,0.26360000)
 rgb(0.75294118)=(0.76270000,0.44010000,0.25480000)
 rgb(0.75686275)=(0.75960000,0.43310000,0.24590000)
 rgb(0.76078431)=(0.75650000,0.42620000,0.23700000)
 rgb(0.76470588)=(0.75330000,0.41920000,0.22820000)
 rgb(0.76862745)=(0.75010000,0.41220000,0.21930000)
 rgb(0.77254902)=(0.74670000,0.40500000,0.21050000)
 rgb(0.77647059)=(0.74320000,0.39780000,0.20160000)
 rgb(0.78039216)=(0.73970000,0.39050000,0.19270000)
 rgb(0.78431373)=(0.73590000,0.38310000,0.18390000)
 rgb(0.78823529)=(0.73200000,0.37550000,0.17500000)
 rgb(0.79215686)=(0.72790000,0.36770000,0.16600000)
 rgb(0.79607843)=(0.72350000,0.35990000,0.15710000)
 rgb(0.80000000)=(0.71890000,0.35180000,0.14820000)
 rgb(0.80392157)=(0.71400000,0.34350000,0.13930000)
 rgb(0.80784314)=(0.70880000,0.33500000,0.13050000)
 rgb(0.81176471)=(0.70330000,0.32640000,0.12150000)
 rgb(0.81568627)=(0.69740000,0.31750000,0.11280000)
 rgb(0.81960784)=(0.69120000,0.30850000,0.10410000)
 rgb(0.82352941)=(0.68470000,0.29930000,0.09560000)
 rgb(0.82745098)=(0.67770000,0.28990000,0.08740000)
 rgb(0.83137255)=(0.67050000,0.28050000,0.07920000)
 rgb(0.83529412)=(0.66290000,0.27100000,0.07150000)
 rgb(0.83921569)=(0.65500000,0.26150000,0.06410000)
 rgb(0.84313725)=(0.64700000,0.25210000,0.05710000)
 rgb(0.84705882)=(0.63870000,0.24270000,0.05060000)
 rgb(0.85098039)=(0.63030000,0.23350000,0.04480000)
 rgb(0.85490196)=(0.62170000,0.22440000,0.03940000)
 rgb(0.85882353)=(0.61310000,0.21570000,0.03480000)
 rgb(0.86274510)=(0.60450000,0.20710000,0.03110000)
 rgb(0.86666667)=(0.59590000,0.19870000,0.02820000)
 rgb(0.87058824)=(0.58740000,0.19070000,0.02600000)
 rgb(0.87450980)=(0.57890000,0.18290000,0.02440000)
 rgb(0.87843137)=(0.57050000,0.17540000,0.02330000)
 rgb(0.88235294)=(0.56230000,0.16820000,0.02250000)
 rgb(0.88627451)=(0.55410000,0.16120000,0.02210000)
 rgb(0.89019608)=(0.54600000,0.15440000,0.02190000)
 rgb(0.89411765)=(0.53800000,0.14790000,0.02170000)
 rgb(0.89803922)=(0.53020000,0.14150000,0.02170000)
 rgb(0.90196078)=(0.52240000,0.13530000,0.02180000)
 rgb(0.90588235)=(0.51480000,0.12920000,0.02200000)
 rgb(0.90980392)=(0.50720000,0.12330000,0.02220000)
 rgb(0.91372549)=(0.49970000,0.11750000,0.02250000)
 rgb(0.91764706)=(0.49230000,0.11180000,0.02280000)
 rgb(0.92156863)=(0.48500000,0.10620000,0.02310000)
 rgb(0.92549020)=(0.47780000,0.10060000,0.02350000)
 rgb(0.92941176)=(0.47060000,0.09520000,0.02390000)
 rgb(0.93333333)=(0.46350000,0.08970000,0.02430000)
 rgb(0.93725490)=(0.45650000,0.08430000,0.02480000)
 rgb(0.94117647)=(0.44950000,0.07870000,0.02520000)
 rgb(0.94509804)=(0.44260000,0.07340000,0.02560000)
 rgb(0.94901961)=(0.43570000,0.06790000,0.02610000)
 rgb(0.95294118)=(0.42890000,0.06240000,0.02650000)
 rgb(0.95686275)=(0.42210000,0.05680000,0.02700000)
 rgb(0.96078431)=(0.41540000,0.05110000,0.02740000)
 rgb(0.96470588)=(0.40880000,0.04540000,0.02780000)
 rgb(0.96862745)=(0.40210000,0.03940000,0.02820000)
 rgb(0.97254902)=(0.39560000,0.03340000,0.02860000)
 rgb(0.97647059)=(0.38900000,0.02780000,0.02890000)
 rgb(0.98039216)=(0.38250000,0.02260000,0.02930000)
 rgb(0.98431373)=(0.37600000,0.01760000,0.02960000)
 rgb(0.98823529)=(0.36960000,0.01290000,0.02990000)
 rgb(0.99215686)=(0.36320000,0.00820000,0.03010000)
 rgb(0.99607843)=(0.35680000,0.00400000,0.03030000)
 rgb(1.00000000)=(0.35040000,0.00010000,0.03050000)},
 }

 \pgfplotsset{
 colormap={tableaucolorblind}{rgb(0.00000000)=(0.06666667,0.43921569,0.66666667)
 rgb(0.11111111)=(0.98823529,0.49019608,0.04313725)
 rgb(0.22222222)=(0.63921569,0.67450980,0.72549020)
 rgb(0.33333333)=(0.34117647,0.37647059,0.42352941)
 rgb(0.44444444)=(0.37254902,0.63529412,0.80784314)
 rgb(0.55555556)=(0.78431373,0.32156863,0.00000000)
 rgb(0.66666667)=(0.48235294,0.51764706,0.56078431)
 rgb(0.77777778)=(0.63921569,0.80000000,0.91372549)
 rgb(0.88888889)=(1.00000000,0.73725490,0.47450980)
 rgb(1.00000000)=(0.78431373,0.81568627,0.85098039)},
 }

\pgfplotsset{ layers/my layer set/.define layer set={
        background, backishground, main, foreground
    }{
    },
    set layers=my layer set,
}

\usepackage{tikz}

\usepackage{xspace}
\usepackage{tabularx}
\usepackage{arydshln}
\usepackage[normalem]{ulem}
\usepackage{placeins}
\usepackage[capitalize]{cleveref}
\usepackage{siunitx}
\usepackage{intcalc}
\usepackage[inline]{enumitem}

\makeatletter
\newcommand{\removelatexerror}{\let\@latex@error\@gobble}
\makeatother

\definecolor{DarkGreen}{rgb}{0.1,0.5,0.1}
\definecolor{DarkRed}{rgb}{0.5,0.1,0.1}
\definecolor{DarkBlue}{rgb}{0.1,0.1,0.5}

\newtheorem{corollary}{Corollary}

\newtheorem{definition}{Definition}
\newtheorem{notation}{Notation}

\newtheorem{example}{Example}
\newtheorem{remark}{Remark}
\newtheorem{construction}{Construction}

\newcommand{\defeq}{\vcentcolon=}
\newcommand{\given}{\vert}

\newcommand{\gap}{\operatorname{gap}} %

\newcommand{\vect}{\operatorname{vec}}

\newcommand{\setint}[2]{\{#1 \!:\! #2\}}
\newcommand{\vecint}[2]{[#1 \!:\! #2]}

\newcommand{\wind}{\ensuremath{k\xspace}} %

\newcommand{\cat}{CAT\xspace}
\newcommand{\cats}{CATs\xspace}
\newcommand{\catx}{\text{CAT\textsubscript{x}}\xspace}
\newcommand{\gasp}{\text{GASP}\xspace}
\newcommand{\gaspsmall}{\text{GASP\textsubscript{small}}\xspace}
\newcommand{\gaspbig}{\text{GASP\textsubscript{big}}\xspace}
\newcommand{\gaspr}{\text{GASP\textsubscript{r}}\xspace}

\newcommand{\gasprs}{GASP\textsubscript{rs}\xspace}

\newcommand{\dogrs}{DOG\textsubscript{rs}\xspace}

\newcommand{\ngasprs}{N\textsubscript{GASP\textsubscript{rs}}\xspace}
\newcommand{\ndogrs}{N\textsubscript{DOG\textsubscript{rs}}\xspace}

\newcommand{\polegap}{PoleGap\xspace}

\newcommand{\ncatx}{\ensuremath{N_{\catx}}\xspace}
\newcommand{\ngaspr}{\ensuremath{N_{\gaspr}}\xspace}
\newcommand{\ngaspsmall}{\ensuremath{N_{\gaspsmall}}\xspace}

\newcommand{\MI}{\ensuremath{\mathrm{I}}\xspace}
\newcommand{\En}{\ensuremath{\mathrm{H}}\xspace}

\newcommand{\F}{\ensuremath{\mathbb{F}}\xspace}
\newcommand{\bbF}{\ensuremath{\mathbb{F}}\xspace}
\newcommand{\bbZ}{\ensuremath{\mathbb{Z}}\xspace}
\newcommand{\Zq}{\ensuremath{\mathbb{Z}_q}\xspace}

\newcommand{\bfA}{\ensuremath{\mathbf{A}}\xspace}
\newcommand{\bfAt}{\ensuremath{\widetilde{\mathbf{A}}}\xspace}
\newcommand{\bfB}{\ensuremath{\mathbf{B}}\xspace}
\newcommand{\bfBt}{\ensuremath{\widetilde{\mathbf{B}}}\xspace}

\newcommand{\bfR}{\ensuremath{\mathbf{R}}\xspace}
\newcommand{\bfS}{\ensuremath{\mathbf{S}}\xspace}

\newcommand{\bfF}{\ensuremath{\mathbf{F}}\xspace}
\newcommand{\bfG}{\ensuremath{\mathbf{G}}\xspace}
\newcommand{\bfH}{\ensuremath{\mathbf{H}}\xspace}

\newcommand{\bfV}{\ensuremath{\mathbf{V}}\xspace}
\newcommand{\bfD}{\ensuremath{\mathbf{D}}\xspace}

\newcommand{\bfx}{\ensuremath{\mathbf{x}}\xspace}
\newcommand{\bfy}{\ensuremath{\mathbf{y}}\xspace}

\newcommand{\cT}{\ensuremath{\mathcal{T}}\xspace}
\newcommand{\cA}{\ensuremath{\mathcal{A}}\xspace}
\newcommand{\cB}{\ensuremath{\mathcal{B}}\xspace}

\newcommand{\cX}{\ensuremath{\mathcal{X}}\xspace}

\newcommand{\cI}{\ensuremath{\mathcal{I}}\xspace}

\newcommand{\rvA}{\ensuremath{\mathbf{A}}\xspace}
\newcommand{\rvB}{\ensuremath{\mathbf{B}}\xspace}

\newcommand{\rvAtT}{\ensuremath{\mathbf{\widetilde{A}}^{(\cT)}}\xspace}

\newcommand{\rvBtT}{\ensuremath{\mathbf{\widetilde{B}}^{(\cT)}}\xspace}
\newcommand{\bfAtT}{\ensuremath{\mathbf{\widetilde{A}}^{(\cT)}}\xspace}
\newcommand{\bfBtT}{\ensuremath{\mathbf{\widetilde{B}}^{(\cT)}}\xspace}

\newcommand{\Ks}{\ensuremath{{K^\star}}\xspace}
\newcommand{\Ls}{\ensuremath{{L^\star}}\xspace}
\newcommand{\Tm}{\ensuremath{\bar{T}}\xspace}

\newcommand{\TL}{\ensuremath{\mathcal{TL}}\xspace}
\newcommand{\TR}{\ensuremath{\mathcal{TR}}\xspace}
\newcommand{\BL}{\ensuremath{\mathcal{BL}}\xspace}
\newcommand{\BR}{\ensuremath{\mathcal{BR}}\xspace}

\newcommand{\ap}{\ensuremath{\boldsymbol{\alpha}^{(p)}}\xspace}
\newcommand{\bp}{\ensuremath{\boldsymbol{\beta}^{(p)}}\xspace}
\newcommand{\as}{\ensuremath{\boldsymbol{\alpha}^{(s)}}\xspace}
\newcommand{\bs}{\ensuremath{\boldsymbol{\beta}^{(s)}}\xspace}

\newcommand{\asind}[1]{\ensuremath{\alpha_{#1}^{(s)}}\xspace}

\newcommand{\sol}[1]{\ensuremath{\mathbf{z}_{#1}}\xspace}
\newcommand{\rect}{\ensuremath{\mathcal{R}}\xspace}

\newcommand{\lat}{\ensuremath{\mathcal{L}}\xspace}

\newcommand{\nrowA}{\ensuremath{{r_{A}}}\xspace}

\newcommand{\ncolA}{\ensuremath{{c_{A}}}\xspace}
\newcommand{\nrowB}{\ensuremath{{r_{B}}}\xspace}
\newcommand{\ncolB}{\ensuremath{{c_{B}}}\xspace}

\begin{document}

\title{CAT and DOG: Improved Codes for Private \\Distributed Matrix Multiplication}

\author{
   \IEEEauthorblockN{Christoph Hofmeister, Rawad Bitar, and Antonia Wachter-Zeh}\\
  \IEEEauthorblockA{Technical University of Munich (TUM), \{christoph.hofmeister, rawad.bitar, antonia.wachter-zeh\}@tum.de}
  \thanks{
    This project has received funding from the German Research Foundation (DFG) under Grant Agreement Nos. WA 3907/7-1 and BI 2492/1-1.
  The authors thank Emma Munisamy and Yonatan Yehezkeally for insightful discussions.
   \ifarxiv \else  A preprint including supplementary material is available online~\cite{arxivversion}. \\ \phantom{.} \fi }
\vspace{-3ex}
}

\ifarxiv
  \markboth{Submitted to the 2025 IEEE International Symposium on Information Theory}{}
\else
    \pagestyle{empty}
\fi

\maketitle

\ifarxiv \else \thispagestyle{empty} \fi
    
\begin{abstract}
We present novel constructions of polynomial codes for private distributed matrix multiplication (PDMM/SDMM) using outer product partitioning (OPP). 
We extend the degree table framework from the literature to \emph{cyclic-addition degree tables} (CATs). 
By using roots of unity as evaluation points, we enable modulo-addition in the table. 
Based on CATs, we present an explicit construction, called \catx, that requires fewer workers than existing schemes in the low-privacy regime. 
Additionally, we present new families of schemes based on conventional degree tables, called \gasprs and \dogrs, that outperform the state-of-the-art for a wide range of parameters.
\end{abstract}

\section{Introduction} \label{sec:intro}
The multiplication of two matrices $\bfA$ and $\bfB$ shall be distributed from a main node to a number $N$ of worker nodes, any $T$ of which shall gain no information about the values of $\bfA$ and $\bfB$. The goal is to design a scheme that requires as few workers as possible while guaranteeing a fixed privacy parameter $T$ and successful decoding of $\bfA \bfB$ at the main node.

This problem, called \emph{private (or secure) distributed matrix multiplication (PDMM/SDMM)}, has received considerable attention %
\cite{chang2018capacity,byrne2023straggler,li2022efficient,lopez2022secure,kakar2019capacity,d2020notes,machado2023hera,makkonen2024flexible,makkonen2024general,hasircioglu2022bivariate,cartor2024secure}.
The PDMM problem is part of the wider literature on coding for privacy, straggler-tolerance and resilience to malicious workers in linear and polynomial computations~\cite{lee2018speeding, bitar2020minimizing,bitar2021private, gomez-vilardebo2024generalized,yu2019lagrange,
yu2021coded,jia2021capacity,morteza2024distributed,censor2024near,malak2024structured,raviv2020private,kiani2022successive,bitar2024sparsity}. 
Commonly in PDMM, $\bfA$ and $\bfB$ are partitioned into smaller blocks and encoded into tasks sent to the workers. We focus on the so-called outer product partitioning (OPP), where $\bfA$ is split horizontally into $K$ equally\footnote{The sizes of \bfA and \bfB may need to be slightly increased by zero-padding.} sized blocks, i.e., $\bfA = \left(\bfA_1^T~\dots~\bfA_K^T\right)^T$ and $\bfB$ vertically into $L$ equally sized blocks, i.e., $\bfB = \left(\bfB_1~\dots~\bfB_L\right)$.
Typically,\footnote{This includes all schemes for the OPP, that we are aware of.} for fixed $K$, $L$ and $T$ the amount of computation performed at each worker is fixed, while the %
upload and download costs are proportional to $N$.
Thus, a scheme is favorable over another if it uses fewer workers.

For the OPP, \gaspr~\cite{doliveira2021degreec} and \polegap~\cite{makkonen2023algebraica} require the lowest numbers of workers among the schemes in the literature.

\emph{Contributions:} %
\begin{enumerate*}
  \item We extend the degree table framework of~\cite{doliveira2020GASPa,doliveira2021degreec} to cyclic-addition degree tables (CATs), by restricting the evaluation points to certain roots of unity and prove that they correspond to %
  PDMM schemes. 
  \item Based on the \cat framework, we design the scheme \catx, that outperforms existing solutions in the low privacy regime $K,L >> T$. %
  \item We present new constructions, \gasprs and \dogrs, for the original degree table framework that use fewer workers than existing solutions for many parameters.
\end{enumerate*}

\Cref{fig:bestschemeKisL} illustrates a range of parameters where the new schemes, \catx and \dogrs, use fewer workers than the state-of-the-art.

\begin{figure}[t]
    \centering\resizebox{0.65\linewidth}{!}{\input{tikz/bestscheme_K=L}}
    \vspace{-0.3cm}
    \caption{\small 
      The color indicates which scheme uses fewest workers. %
    The numbers show the difference to the second best scheme.
  }
    \label{fig:bestschemeKisL}
    \vspace{-0.3cm}
\end{figure}

\emph{Related Work:} The main novelty in \cat over degree tables \cite{doliveira2021degreec} is the use of roots of unity as evaluation points, enabling modulo-addition in the table. These and related concepts have been previously used in the literature on PDMM. For instance, \cite{machado2022root,karpuk2024modular,makkonen2022analog} use roots of unity as evaluation points, \cite{mital2022secure} uses the discrete Fourier transform, and \cite{aliasgari2020private} uses cyclic convolutions. However, previous works do not apply these concepts in a way that achieves fewer workers than \gaspr and \polegap in OPP.

\emph{Notation:}
Let $\bbF_p$ denote the field with $p$ elements and $\bbZ_q$ denote the %
integers modulo $q$. %
For integers $a$ and $b$, define the vector $\vecint{a}{b} \defeq (a, a+1, \dots, b)$ and the set $\setint{a}{b} \defeq \{a, a+1, \dots, b\}$. Coprime integers are denoted as $a \perp b$.
For vectors $\bfx$ and $\bfy$, $\bfx || \bfy$ denotes their concatenation. For sets $\cA$ and $\cB$, define the sumset $\cA+\cB\defeq\{a+b | a\in \cA, b\in \cB\}$. Further, let $a+\cB\defeq \{a+b|b\in \cB\}$ and $a\cdot\cB\defeq\{a\cdot b | b \in \cB\}$.
For a vector $\bfx= (x_1, \dots, x_{n_x})$ of length $n_x$, let $\{\bfx\} \defeq \{x_i | i \in \setint{1}{n_x}\}$ and conversely for a set $\cX$ of integers, let $\vect(\cX)$ denote the vector containing the elements of $\cX$ in ascending order.
Let $a+ \mathbf{x} \defeq (a+x_1,\dots, a+x_{n_x})$.
For vectors $\bfx$ of length $n_x$ and $\bfy$ of length $n_y$, let $\bfV(\bfx, \bfy)$ denote the generalized Vandermonde matrix $\bfV(\bfx, \bfy) \defeq (x_i^{y_j})_{1\leq i \leq n_x, 1\leq j \leq n_y}$, i.e., the $n_x \times n_y$ matrix with $(i,j)$th entry equal to $x_i^{y_j}$. %

\section{Motivating Example: $K=L=T=2$} \label{sec:introexample}

Let $\bfA$ and $\bfB$ be drawn from an unknown joint distribution over $\bbF^{\nrowA \times \ncolA}_{p=11}$ and $\bbF_{p=11}^{\nrowB \times \ncolB}$, where $\nrowA, \ncolA=\nrowB, \ncolB$ are natural numbers and $\nrowA$ and $\ncolB$ are even. 
The matrices are partitioned into %
$\bfA  = \begin{pmatrix} \bfA_1^T & \bfA_2^T \end{pmatrix}^T$ and $\bfB=\begin{pmatrix} \bfB_1 & \bfB_2 \end{pmatrix}$.
The goal is for the main node to obtain
        $\bfA_1 \bfB_1$, $\bfA_1 \bfB_2$,
        $\bfA_2 \bfB_1$, and $\bfA_2 \bfB_2$ while protecting the privacy of $\bfA$ and $\bfB$, and using as few workers as possible. The data sent by the main node to any $T=2$ of the workers has to be statistically independent of $\bfA$ and $\bfB$. For $K=L=T=2$, \gaspr and \polegap require $N=11$ workers. Using \cat, we achieve $N=10$.

      \emph{Encoding:} To fulfill the privacy requirement, the main node encodes $\bfA$ and $\bfB$ into \emph{tasks} for the workers, using random matrices $\bfR_1,\bfR_2 \in \bbF_{11}^{\frac{\nrowA}{2} \times \ncolA}$ and $\bfS_1,\bfS_2 \in \bbF_{11}^{\nrowB \times \frac{\ncolB}{2}}$. These matrices are drawn independently and uniformly at random.
The encoding is done via the following polynomials and forms a secret sharing scheme~\cite{shamir1979share,mceliece1981sharing}
\begin{align*}
        \bfF(x) &= \bfA_1 + \bfA_2 x^3 + \bfR_1 x^6 + \bfR_2 x^7, \\ 
        \bfG(x) &= \bfB_1 + \bfB_2 x + \bfS_1 x^9 + \bfS_2 x^2,
\end{align*}
where the degrees are carefully selected together with a \emph{modulus} $q\geq N$. In this case, we choose $q=10$. 
The main node picks a distinct $(q=10)$th \emph{root-of-unity} from $\bbF_{11}$ for each worker. 
To this end, it selects an element $\omega$ of order $10$, specifically $\omega=2$, and computes the consecutive powers $\boldsymbol{\rho} \defeq (\omega^0, \omega^1, \dots, \omega^9) = (1, 2, 4, 8, 5, 10, 9, 7, 3, 6)$.
The task sent to each worker $\wind \in \setint{1}{N}$ consists of the evaluations $\bfF(\rho_\wind)$ and $\bfG(\rho_\wind)$. The worker computes the product and returns $\bfF(\rho_\wind)\bfG(\rho_\wind)$.

\emph{Decoding:}
Define $\bfH(x) \defeq \bfF(x)\bfG(x) \bmod (x^{10}-1)$. 
Reducing a polynomial modulo $x^q-1$ corresponds to reducing all degrees modulo $q$, i.e., $x^q \equiv x^0 \pmod{x^q-1}$.
Since the workers' evaluations are at $10$th roots of unity (i.e. $\rho_\wind^{10}=\rho_\wind^0=1$), we can write $\bfF(\omega^\wind)\bfG(\omega^\wind) = \bfH(\omega^\wind)$, $\wind\in\setint{1}{N}$. 
The degrees of the monomials in $\bfF(x)$ and $\bfG(x)$ can be related to the degrees of the monomials in $\bfH(x)$ using an addition table (over the cyclic group $\bbZ_{q=10}$) as shown in \cref{fig:introcat222}. 
  \begin{figure}
    \begin{subfigure}{0.48\linewidth}
        \centering
        \resizebox{0.7\linewidth}{!}{
        \begin{tikzpicture}[
 /tikz/background rectangle/.style={fill={rgb,1:red,1.0;green,1.0;blue,1.0}, fill opacity={1.0}, draw opacity={1.0}}, show background rectangle]
\begin{axis}[point meta max={9.0}, point meta min={0.0}, legend cell align={left}, legend columns={1}, title={}, title style={at={{(0.5,1)}}, anchor={south}, font={{\fontsize{14 pt}{18.2 pt}\selectfont}}, color={rgb,1:red,0.0;green,0.0;blue,0.0}, draw opacity={1.0}, rotate={0.0}, align={center}}, legend style={color={rgb,1:red,0.0;green,0.0;blue,0.0}, draw opacity={1.0}, line width={1}, solid, fill={rgb,1:red,1.0;green,1.0;blue,1.0}, fill opacity={1.0}, text opacity={1.0}, font={{\fontsize{8 pt}{10.4 pt}\selectfont}}, text={rgb,1:red,0.0;green,0.0;blue,0.0}, cells={anchor={center}}, at={(1.02, 1)}, anchor={north west}}, axis background/.style={fill={rgb,1:red,1.0;green,1.0;blue,1.0}, opacity={1.0}}, anchor={north west}, xshift={1.0mm}, yshift={-1.0mm}, 
  width={100mm}, height={100mm}, 
  scaled x ticks={false}, xlabel={}, x tick style={color={rgb,1:red,0.0;green,0.0;blue,0.0}, opacity={1.0}}, x tick label style={color={rgb,1:red,0.0;green,0.0;blue,0.0}, opacity={1.0}, rotate={0}}, xlabel style={at={(ticklabel cs:0.5)}, anchor=near ticklabel, at={{(ticklabel cs:0.5)}}, anchor={near ticklabel}, font={{\fontsize{7 pt}{14.3 pt}\selectfont}}, color={rgb,1:red,0.0;green,0.0;blue,0.0}, draw opacity={1.0}, rotate={0.0}}, xmajorgrids={false}, xmin={0.5}, xmax={4.5}, xticklabels={{$\bfB_1$,$\bfB_2x$,$\bfS_1x^9$,$\bfS_2x^2$}}, xtick={{1,2,3,4}}, xtick align={outside}, xticklabel style={font={{\fontsize{46 pt}{59.800000000000004 pt}\selectfont}}, color={rgb,1:red,0.0;green,0.0;blue,0.0}, draw opacity={1.0}, rotate={0.0}}, x grid style={color={rgb,1:red,0.0;green,0.0;blue,0.0}, draw opacity={0.1}, line width={0.5}, solid}, axis x line*={right}, x axis line style={color={rgb,1:red,0.0;green,0.0;blue,0.0}, draw opacity={1.0}, line width={1}, solid}, scaled y ticks={false}, ylabel={}, y tick style={color={rgb,1:red,0.0;green,0.0;blue,0.0}, opacity={1.0}}, y tick label style={color={rgb,1:red,0.0;green,0.0;blue,0.0}, opacity={1.0}, rotate={0}}, ylabel style={at={(ticklabel cs:0.5)}, anchor=near ticklabel, at={{(ticklabel cs:0.5)}}, anchor={near ticklabel}, font={{\fontsize{8 pt}{14.3 pt}\selectfont}}, color={rgb,1:red,0.0;green,0.0;blue,0.0}, draw opacity={1.0}, rotate={0.0}}, y dir={reverse}, ymajorgrids={false}, ymin={0.5}, ymax={4.5}, yticklabels={{$\bfA_1$,$\bfA_2x^3$,$\bfR_1x^6$,$\bfR_2x^7$}}, ytick={{1,2,3,4}}, ytick align={outside}, yticklabel style={font={{\fontsize{46 pt}{59.800000000000004 pt}\selectfont}}, color={rgb,1:red,0.0;green,0.0;blue,0.0}, draw opacity={1.0}, rotate={0.0}}, y grid style={color={rgb,1:red,0.0;green,0.0;blue,0.0}, draw opacity={0.1}, line width={0.5}, solid}, axis y line*={left}, y axis line style={color={rgb,1:red,0.0;green,0.0;blue,0.0}, draw opacity={1.0}, line width={1}, solid}, colormap name={plots1}, colorbar={false}, view={{0}{90}}]
    \addplot3[color={rgb,1:red,0.8321;green,0.7477;blue,0.7052}, name path={73}, matrix plot*, mesh/rows={4}, mesh/cols={4}, point meta={\thisrow{meta}}, opacity={1.0}, forget plot]
        table[row sep={\\}]
        {
            x  y  z  meta  \\
            1.0  1.0  0.0  0.0  \\
            2.0  1.0  1.0  1.0  \\
            3.0  1.0  9.0  9.0  \\
            4.0  1.0  2.0  2.0  \\
            1.0  2.0  3.0  3.0  \\
            2.0  2.0  4.0  4.0  \\
            3.0  2.0  2.0  2.0  \\
            4.0  2.0  5.0  5.0  \\
            1.0  3.0  6.0  6.0  \\
            2.0  3.0  7.0  7.0  \\
            3.0  3.0  5.0  5.0  \\
            4.0  3.0  8.0  8.0  \\
            1.0  4.0  7.0  7.0  \\
            2.0  4.0  8.0  8.0  \\
            3.0  4.0  6.0  6.0  \\
            4.0  4.0  9.0  9.0  \\
        }
        ;
    \addplot[color={rgb,1:red,0.0;green,0.0;blue,0.0}, name path={74}, area legend, fill={none}, fill opacity={1.0}, draw opacity={1.0}, line width={3}, solid, forget plot]
        table[row sep={\\}]
        {
            \\
            0.5  0.5  \\
            2.5  0.5  \\
            2.5  2.5  \\
            0.5  2.5  \\
            0.5  0.5  \\
        }
        ;
    \addplot[color={rgb,1:red,0.0;green,0.0;blue,0.0}, name path={75}, area legend, fill={none}, fill opacity={1.0}, draw opacity={1.0}, line width={3}, solid, forget plot]
        table[row sep={\\}]
        {
            \\
            0.5  0.5  \\
            4.5  0.5  \\
            4.5  4.5  \\
            0.5  4.5  \\
            0.5  0.5  \\
        }
        ;
    \addplot[color={rgb,1:red,0.0;green,0.0;blue,0.0}, name path={76}, area legend, fill={none}, fill opacity={1.0}, draw opacity={1.0}, line width={3}, solid, forget plot]
        table[row sep={\\}]
        {
            \\
            2.5  2.5  \\
            4.5  2.5  \\
            4.5  4.5  \\
            2.5  4.5  \\
            2.5  2.5  \\
        }
        ;
    \node[, color={rgb,1:red,1.0;green,1.0;blue,1.0}, draw opacity={1.0}, rotate={0.0}, font={{\fontsize{35 pt}{45.5 pt}\selectfont}}]  at (axis cs:1,1) {0};
    \node[, color={rgb,1:red,1.0;green,1.0;blue,1.0}, draw opacity={1.0}, rotate={0.0}, font={{\fontsize{35 pt}{45.5 pt}\selectfont}}]  at (axis cs:2,1) {1};
    \node[, color={rgb,1:red,1.0;green,1.0;blue,1.0}, draw opacity={1.0}, rotate={0.0}, font={{\fontsize{35 pt}{45.5 pt}\selectfont}}]  at (axis cs:3,1) {9};
    \node[, color={rgb,1:red,1.0;green,1.0;blue,1.0}, draw opacity={1.0}, rotate={0.0}, font={{\fontsize{35 pt}{45.5 pt}\selectfont}}]  at (axis cs:4,1) {2};
    \node[, color={rgb,1:red,1.0;green,1.0;blue,1.0}, draw opacity={1.0}, rotate={0.0}, font={{\fontsize{35 pt}{45.5 pt}\selectfont}}]  at (axis cs:1,2) {3};
    \node[, color={rgb,1:red,1.0;green,1.0;blue,1.0}, draw opacity={1.0}, rotate={0.0}, font={{\fontsize{35 pt}{45.5 pt}\selectfont}}]  at (axis cs:2,2) {4};
    \node[, color={rgb,1:red,1.0;green,1.0;blue,1.0}, draw opacity={1.0}, rotate={0.0}, font={{\fontsize{35 pt}{45.5 pt}\selectfont}}]  at (axis cs:3,2) {2};
    \node[, color={rgb,1:red,1.0;green,1.0;blue,1.0}, draw opacity={1.0}, rotate={0.0}, font={{\fontsize{35 pt}{45.5 pt}\selectfont}}]  at (axis cs:4,2) {5};
    \node[, color={rgb,1:red,1.0;green,1.0;blue,1.0}, draw opacity={1.0}, rotate={0.0}, font={{\fontsize{35 pt}{45.5 pt}\selectfont}}]  at (axis cs:1,3) {6};
    \node[, color={rgb,1:red,1.0;green,1.0;blue,1.0}, draw opacity={1.0}, rotate={0.0}, font={{\fontsize{35 pt}{45.5 pt}\selectfont}}]  at (axis cs:2,3) {7};
    \node[, color={rgb,1:red,1.0;green,1.0;blue,1.0}, draw opacity={1.0}, rotate={0.0}, font={{\fontsize{35 pt}{45.5 pt}\selectfont}}]  at (axis cs:3,3) {5};
    \node[, color={rgb,1:red,1.0;green,1.0;blue,1.0}, draw opacity={1.0}, rotate={0.0}, font={{\fontsize{35 pt}{45.5 pt}\selectfont}}]  at (axis cs:4,3) {8};
    \node[, color={rgb,1:red,1.0;green,1.0;blue,1.0}, draw opacity={1.0}, rotate={0.0}, font={{\fontsize{35 pt}{45.5 pt}\selectfont}}]  at (axis cs:1,4) {7};
    \node[, color={rgb,1:red,1.0;green,1.0;blue,1.0}, draw opacity={1.0}, rotate={0.0}, font={{\fontsize{35 pt}{45.5 pt}\selectfont}}]  at (axis cs:2,4) {8};
    \node[, color={rgb,1:red,1.0;green,1.0;blue,1.0}, draw opacity={1.0}, rotate={0.0}, font={{\fontsize{35 pt}{45.5 pt}\selectfont}}]  at (axis cs:3,4) {6};
    \node[, color={rgb,1:red,1.0;green,1.0;blue,1.0}, draw opacity={1.0}, rotate={0.0}, font={{\fontsize{35 pt}{45.5 pt}\selectfont}}]  at (axis cs:4,4) {9};
\end{axis}
\end{tikzpicture}}
        \caption{\small Cyclic-addition degree table. \\ \phantom{(a)} (Modulo-$10$ addition).}
        \label{fig:introcat222}
      \end{subfigure}
      \begin{subfigure}{0.48\linewidth}
        \centering
        \resizebox{0.7\linewidth}{!}{
        \begin{tikzpicture}[
 /tikz/background rectangle/.style={fill={rgb,1:red,1.0;green,1.0;blue,1.0}, fill opacity={1.0}, draw opacity={1.0}}, show background rectangle]
\begin{axis}[point meta max={11.0}, point meta min={0.0}, legend cell align={left}, legend columns={1}, title={}, title style={at={{(0.5,1)}}, anchor={south}, font={{\fontsize{14 pt}{18.2 pt}\selectfont}}, color={rgb,1:red,0.0;green,0.0;blue,0.0}, draw opacity={1.0}, rotate={0.0}, align={center}}, legend style={color={rgb,1:red,0.0;green,0.0;blue,0.0}, draw opacity={1.0}, line width={1}, solid, fill={rgb,1:red,1.0;green,1.0;blue,1.0}, fill opacity={1.0}, text opacity={1.0}, font={{\fontsize{8 pt}{10.4 pt}\selectfont}}, text={rgb,1:red,0.0;green,0.0;blue,0.0}, cells={anchor={center}}, at={(1.02, 1)}, anchor={north west}}, axis background/.style={fill={rgb,1:red,1.0;green,1.0;blue,1.0}, opacity={1.0}}, anchor={north west}, xshift={1.0mm}, yshift={-1.0mm}, 
  width={100mm}, height={100mm}, 
  scaled x ticks={false}, xlabel={}, x tick style={color={rgb,1:red,0.0;green,0.0;blue,0.0}, opacity={1.0}}, x tick label style={color={rgb,1:red,0.0;green,0.0;blue,0.0}, opacity={1.0}, rotate={0}}, xlabel style={at={(ticklabel cs:0.5)}, anchor=near ticklabel, at={{(ticklabel cs:0.5)}}, anchor={near ticklabel}, font={{\fontsize{7 pt}{14.3 pt}\selectfont}}, color={rgb,1:red,0.0;green,0.0;blue,0.0}, draw opacity={1.0}, rotate={0.0}}, xmajorgrids={false}, xmin={0.5}, xmax={4.5}, xticklabels={{$\bfB_1$,$\bfB_2x^2$,$\bfS_1x^4$,$\bfS_2x^5$}}, xtick={{1,2,3,4}}, xtick align={outside}, xticklabel style={font={{\fontsize{46 pt}{59.800000000000004 pt}\selectfont}}, color={rgb,1:red,0.0;green,0.0;blue,0.0}, draw opacity={1.0}, rotate={0.0}}, x grid style={color={rgb,1:red,0.0;green,0.0;blue,0.0}, draw opacity={0.1}, line width={0.5}, solid}, axis x line*={right}, x axis line style={color={rgb,1:red,0.0;green,0.0;blue,0.0}, draw opacity={1.0}, line width={1}, solid}, scaled y ticks={false}, ylabel={}, y tick style={color={rgb,1:red,0.0;green,0.0;blue,0.0}, opacity={1.0}}, y tick label style={color={rgb,1:red,0.0;green,0.0;blue,0.0}, opacity={1.0}, rotate={0}}, ylabel style={at={(ticklabel cs:0.5)}, anchor=near ticklabel, at={{(ticklabel cs:0.5)}}, anchor={near ticklabel}, font={{\fontsize{8 pt}{14.3 pt}\selectfont}}, color={rgb,1:red,0.0;green,0.0;blue,0.0}, draw opacity={1.0}, rotate={0.0}}, y dir={reverse}, ymajorgrids={false}, ymin={0.5}, ymax={4.5}, yticklabels={{$\bfA_1$,$\bfA_2x$,$\bfR_1x^4$,$\bfR_2x^6$}}, ytick={{1,2,3,4}}, ytick align={outside}, yticklabel style={font={{\fontsize{46 pt}{59.800000000000004 pt}\selectfont}}, color={rgb,1:red,0.0;green,0.0;blue,0.0}, draw opacity={1.0}, rotate={0.0}}, y grid style={color={rgb,1:red,0.0;green,0.0;blue,0.0}, draw opacity={0.1}, line width={0.5}, solid}, axis y line*={left}, y axis line style={color={rgb,1:red,0.0;green,0.0;blue,0.0}, draw opacity={1.0}, line width={1}, solid}, colormap name={plots1}, colorbar={false}, view={{0}{90}}]
    \addplot3[color={rgb,1:red,0.8321;green,0.7477;blue,0.7052}, name path={73}, matrix plot*, mesh/rows={4}, mesh/cols={4}, point meta={\thisrow{meta}}, opacity={1.0}, forget plot]
        table[row sep={\\}]
        {
            x  y  z  meta  \\
            1.0  1.0  0.0  0.0  \\
            2.0  1.0  2.0  2.0  \\
            3.0  1.0  4.0  4.0  \\
            4.0  1.0  5.0  5.0  \\
            1.0  2.0  1.0  1.0  \\
            2.0  2.0  3.0  3.0  \\
            3.0  2.0  5.0  5.0  \\
            4.0  2.0  6.0  6.0  \\
            1.0  3.0  4.0  4.0  \\
            2.0  3.0  6.0  6.0  \\
            3.0  3.0  8.0  8.0  \\
            4.0  3.0  9.0  9.0  \\
            1.0  4.0  6.0  6.0  \\
            2.0  4.0  8.0  8.0  \\
            3.0  4.0  10.0  10.0  \\
            4.0  4.0  11.0  11.0  \\
        }
        ;
    \addplot[color={rgb,1:red,0.0;green,0.0;blue,0.0}, name path={4d860a41-5ff6-4ba9-a857-13fc630e482e}, area legend, fill={none}, fill opacity={1.0}, draw opacity={1.0}, line width={3}, solid, forget plot]
        table[row sep={\\}]
        {
            \\
            0.5  0.5  \\
            2.5  0.5  \\
            2.5  2.5  \\
            0.5  2.5  \\
            0.5  0.5  \\
        }
        ;
    \addplot[color={rgb,1:red,0.0;green,0.0;blue,0.0}, name path={0013e47c-0aa4-4b61-8253-55d71e5b7d1b}, area legend, fill={none}, fill opacity={1.0}, draw opacity={1.0}, line width={3}, solid, forget plot]
        table[row sep={\\}]
        {
            \\
            0.5  0.5  \\
            4.5  0.5  \\
            4.5  4.5  \\
            0.5  4.5  \\
            0.5  0.5  \\
        }
        ;
    \addplot[color={rgb,1:red,0.0;green,0.0;blue,0.0}, name path={cd4228ea-11b7-462b-8289-de94483dd68c}, area legend, fill={none}, fill opacity={1.0}, draw opacity={1.0}, line width={3}, solid, forget plot]
        table[row sep={\\}]
        {
            \\
            2.5  2.5  \\
            4.5  2.5  \\
            4.5  4.5  \\
            2.5  4.5  \\
            2.5  2.5  \\
        }
        ;
    \node[, color={rgb,1:red,1.0;green,1.0;blue,1.0}, draw opacity={1.0}, rotate={0.0}, font={{\fontsize{35 pt}{45.5 pt}\selectfont}}]  at (axis cs:1,1) {0};
    \node[, color={rgb,1:red,1.0;green,1.0;blue,1.0}, draw opacity={1.0}, rotate={0.0}, font={{\fontsize{35 pt}{45.5 pt}\selectfont}}]  at (axis cs:2,1) {2};
    \node[, color={rgb,1:red,1.0;green,1.0;blue,1.0}, draw opacity={1.0}, rotate={0.0}, font={{\fontsize{35 pt}{45.5 pt}\selectfont}}]  at (axis cs:3,1) {4};
    \node[, color={rgb,1:red,1.0;green,1.0;blue,1.0}, draw opacity={1.0}, rotate={0.0}, font={{\fontsize{35 pt}{45.5 pt}\selectfont}}]  at (axis cs:4,1) {5};
    \node[, color={rgb,1:red,1.0;green,1.0;blue,1.0}, draw opacity={1.0}, rotate={0.0}, font={{\fontsize{35 pt}{45.5 pt}\selectfont}}]  at (axis cs:1,2) {1};
    \node[, color={rgb,1:red,1.0;green,1.0;blue,1.0}, draw opacity={1.0}, rotate={0.0}, font={{\fontsize{35 pt}{45.5 pt}\selectfont}}]  at (axis cs:2,2) {3};
    \node[, color={rgb,1:red,1.0;green,1.0;blue,1.0}, draw opacity={1.0}, rotate={0.0}, font={{\fontsize{35 pt}{45.5 pt}\selectfont}}]  at (axis cs:3,2) {5};
    \node[, color={rgb,1:red,1.0;green,1.0;blue,1.0}, draw opacity={1.0}, rotate={0.0}, font={{\fontsize{35 pt}{45.5 pt}\selectfont}}]  at (axis cs:4,2) {6};
    \node[, color={rgb,1:red,1.0;green,1.0;blue,1.0}, draw opacity={1.0}, rotate={0.0}, font={{\fontsize{35 pt}{45.5 pt}\selectfont}}]  at (axis cs:1,3) {4};
    \node[, color={rgb,1:red,1.0;green,1.0;blue,1.0}, draw opacity={1.0}, rotate={0.0}, font={{\fontsize{35 pt}{45.5 pt}\selectfont}}]  at (axis cs:2,3) {6};
    \node[, color={rgb,1:red,1.0;green,1.0;blue,1.0}, draw opacity={1.0}, rotate={0.0}, font={{\fontsize{35 pt}{45.5 pt}\selectfont}}]  at (axis cs:3,3) {8};
    \node[, color={rgb,1:red,1.0;green,1.0;blue,1.0}, draw opacity={1.0}, rotate={0.0}, font={{\fontsize{35 pt}{45.5 pt}\selectfont}}]  at (axis cs:4,3) {9};
    \node[, color={rgb,1:red,1.0;green,1.0;blue,1.0}, draw opacity={1.0}, rotate={0.0}, font={{\fontsize{35 pt}{45.5 pt}\selectfont}}]  at (axis cs:1,4) {6};
    \node[, color={rgb,1:red,1.0;green,1.0;blue,1.0}, draw opacity={1.0}, rotate={0.0}, font={{\fontsize{35 pt}{45.5 pt}\selectfont}}]  at (axis cs:2,4) {8};
    \node[, color={rgb,1:red,1.0;green,1.0;blue,1.0}, draw opacity={1.0}, rotate={0.0}, font={{\fontsize{35 pt}{45.5 pt}\selectfont}}]  at (axis cs:3,4) {10};
    \node[, color={rgb,1:red,1.0;green,1.0;blue,1.0}, draw opacity={1.0}, rotate={0.0}, font={{\fontsize{35 pt}{45.5 pt}\selectfont}}]  at (axis cs:4,4) {11};
\end{axis}
\end{tikzpicture}}
        \caption{\small $\gaspr$ degree table. %
        (Integer addition).}
        \label{fig:introGASP222}
      \end{subfigure}
        \caption{\small A cyclic-addition degree table and a \gaspr degree table. Each row corresponds to a monomial in $\bfF(x)$. Each column to a monomial in $\bfG(x)$.
        Each cell corresponds to a monomial in $\bfH(x)$ and contains the sum of the corresponding row and column degrees.}
        \label{fig:intro222}
        \vspace{-1em}
  \end{figure}
      Accordingly, we have
      \begin{align*}
        \bfH(x) =& \bfA_1 \bfB_1 + \bfA_1 \bfB_2 x + 
             (\bfA_1 \bfS_2 + \bfA_2 \bfS_1) x^2 + \\
             & \bfA_2 \bfB_1 x^3 + \bfA_2 \bfB_2 x^4 +
             (\bfA_2\bfS_2 + \bfR_1\bfS_1) x^5 + \\
             &(\bfR_1\bfB_1+ \bfR_2\bfS_1) x^6 +
             (\bfR_1 \bfB_2 + \bfR_2 \bfB_1) x^7  + \\
             &(\bfR_2 \bfB_2 + \bfR_1 \bfS_2)x^8 +
             (\bfA_1 \bfS_1 + \bfR_2 \bfS_2) x^9.
      \end{align*}
    After receiving one evaluation of $\bfH(x)$ from each worker, the main node interpolates $\bfH(x)$, i.e., recovers its coefficients; thus obtaining
    the desired computation result from the coefficients of $x^0$, $x^1$, $x^3$, and $x^4$.

    Interpolating $\bfH(x)$ corresponds to solving a linear system with one variable per non-zero coefficient of $\bfH(x)$ and one equation per worker. This system is determined by the matrix $\bfV(\boldsymbol{\rho}, \boldsymbol{\gamma})$, where  $\boldsymbol{\gamma}$ contains the distinct degrees in $\bfH(x)$ (in ascending order). Thus, (provided $\bfV(\boldsymbol{\rho}, \boldsymbol{\gamma})$ is invertible) the number of required workers of the scheme equals the number of non-zero coefficients in $\bfH(x)$, i.e, the number of distinct entries in \cref{fig:introcat222}. 

    \gaspr does not use roots of unity as evaluation points. There is one more distinct entry in the best \gaspr degree table and thus one more worker is required, cf. \cref{fig:introGASP222}.

\emph{Privacy:} To demonstrate that no two workers can gain information about $\bfA$ and $\bfB$ we show that $\bfA_1$, $\bfA_2$, $\bfB_1$ and $\bfB_2$ are perfectly obscured by the random matrices.
Consider any two workers $\wind_1, \wind_2 \in \setint{0}{9}$ with evaluation points $\omega^{\wind_1}$ and $\omega^{\wind_2}$. Then, 
  $\bfF(\omega^{\wind_1}) = \bfA_1 + \bfA_2 \omega^{3 \wind_1} + \bfR_1 \omega^{6 \wind_1} + \bfR_2 \omega^{7\wind_1}$ and $\bfF(\omega^{\wind_2}) = \bfA_1 + \bfA_2 \omega^{3 \wind_2} + \bfR_1 \omega^{6 \wind_2} + \bfR_2 \omega^{7\wind_2}$.
  The relation between pairs of matrices $(\bfR_1, \bfR_2)$ and $(\bfR_1 \omega^{6 \wind_1} + \bfR_2 \omega^{7\wind_1}, \bfR_1 \omega^{6 \wind_2} + \bfR_2 \omega^{7\wind_2})$ is one-to-one. This can be seen by 
  factoring the corresponding matrix into a diagonal matrix and a Vandermonde matrix with distinct rows as\begin{align} \label{eq:exfactor}
  \begin{pmatrix} \omega^{6\wind_1} & \omega^{7\wind_1} \\ \omega^{6\wind_2} & \omega^{7\wind_2} \end{pmatrix} = 
  \begin{pmatrix}
  \omega^{6\wind_1} & 0 \\ 0 & \omega^{6\wind_2}\end{pmatrix} \begin{pmatrix} \omega^0 & \omega^{\wind_1} \\ \omega^0 & \omega^{\wind_2} \end{pmatrix}.
\end{align}
Hence, $\bfR_1 \omega^{6 \wind_1} + \bfR_2 \omega^{7\wind_1}$  and $\bfR_1 \omega^{6 \wind_2} + \bfR_2 \omega^{7\wind_2}$ are uniformly and independently distributed over $\bbF_{11}^{\frac{\nrowA}{2} \times \ncolA}$ (and independent of $\bfA_1$ and $\bfA_2$). 
For any fixed $\bfA_1$ and $\bfA_2$ any two workers' tasks $\bfF(\omega^{\wind_1})$ and $\bfF(\omega^{\wind_2})$ are thus also independently and uniformly distributed over $\bbF_{11}^{\frac{\nrowA}{2} \times \ncolA}$, i.e., $\bfF(\omega^{\wind_1})$ and $\bfF(\omega^{\wind_2})$ are (jointly) statistically independent of $\bfA$.
Analogously, $\bfG(\omega^{\wind_1})$ and  $\bfG(\omega^{\wind_2})$ are independent of $\bfB$. %

\emph{Field Size:} The only requirement on the field is that it contains an element $\omega \in \bbF_p$ of order $q$. Thus, any field can be used (replacing $\bbF_{11}$) as long as $q|p-1$ and $p$ is a prime power, e.g. $11, 31, 41, 61, 71, 81, 101, 121, 131, 151,181, 191, \dots$ %

\emph{Challenges:}
The selection of evaluation points which ensure privacy and decodability requires some care, as does the choice of degrees that allow for such evaluation points.
The matrix in \eqref{eq:exfactor} can always be factored as shown into a diagonal matrix and a second factor. However, the second factor is not necessarily a Vandermonde matrix and may have repeating rows. 
To illustrate this issue, replace the degrees $6$ and $7$ with $1$ and $6$ and consider $2|\wind_1$ and $\wind_2 = 2\wind_1$, e.g., $\wind_1=2$ and $\wind_2=4$. 
Then, the matrix is factored into $\left(\begin{smallmatrix} \omega^{\wind_1} & 0 \\ 0 & \omega^{\wind_2}\end{smallmatrix}\right) \left(\begin{smallmatrix} \omega^0 & \omega^{5\wind_1} \\ \omega^0 & \omega^{5\wind_2} \end{smallmatrix}\right) = \left(\begin{smallmatrix} \omega^{2} & 0 \\ 0 & \omega^{4}\end{smallmatrix}\right) \left(\begin{smallmatrix} \omega^0 & \omega^{0} \\ \omega^0 & \omega^{0} \end{smallmatrix}\right),$ which shows that it is singular.
Thus, in this case the encoding is not private.

\section{Problem Setting} \label{sec:problemsetting}

\subsection{Private Distributed Matrix Multiplication}
A main node distributes the multiplication of two matrices $\bfA \in \F_p^{\nrowA \times \ncolA}$ and $\bfB \in \F_p^{\nrowB \times \ncolB}$ to $N$ worker nodes, where $\nrowA, \ncolA=\nrowB, \ncolB$ are large integers. 
To this end, the main node sends a task $\bfAt_\wind, \bfBt_\wind$, $\wind\in \setint{1}{N}$ to each of $N$ worker nodes.
The matrices $\bfA$ and $\bfB$ must be kept information theoretically $T$-private from any set $\cT\subset \setint{1}{N}$ of up to $|\cT| = T$ colluding workers.
Formally, we require $\MI(\bfA, \bfB; \bfAtT, \bfBtT) = 0$, where $\bfAtT$ and $\bfBtT$ denote the encoded tasks sent to the workers in $\cT$.
We will consider schemes where $\bfA$ and $\bfB$ are encoded independently, in which case $\MI(\bfA; \bfAtT) = \MI(\bfB; \bfBtT) = 0$ is sufficient.

\subsection{Polynomial Codes for PDMM}
In this work, we design and analyze polynomial codes for the OPP. %
    The goal is to enable the main node to reconstructs all pairwise products $\bfA_i\bfB_j$, $i\in\setint{1}{K}, j\in\setint{1}{L}$. 
    The main node draws matrices $\bfR_1, \dots, \bfR_T \in \mathbb{F}_p^{\frac{\nrowA}{K}\times \ncolA}$ and $\bfS_1, \dots, \bfS_T\in \bbF_p^{\nrowB \times \frac{\ncolB}{L}}$ independently and uniformly at random and independently from $\bfA$ and $\bfB$. 
    While other encodings are possible, e.g. \cite{yu2019lagrange}, we consider an encoding of the form $\bfF(x) = \sum_{i=1}^{K} \bfA_i x^{\alpha^{(p)}_i} + \sum_{i=1}^{T} \bfR_i x^{\alpha^{(s)}_{i}}$ and $\bfG(x) = \sum_{i=1}^{L} \bfB_i x^{\beta^{(p)}_i} + \sum_{i=1}^{T} \bfS_i x^{\beta^{(s)}_{i}}$, where the vectors of degrees%
  \footnote{As in~\cite{doliveira2021degreec}, the superscripts $(p)$ and $(s)$ stand for ``prefix'' and ``suffix.''}
    $\ap \in \mathbb{Z}^{K}, \bp \in \mathbb{Z}^{L}, \as \in \mathbb{Z}^{T},$ and $\bs\in \mathbb{Z}^T$ as well as the vector of evaluation points $\boldsymbol{\rho} \in \mathbb{F}_p^{N}$ need to be carefully chosen. 
The main node constructs the tasks for each worker $\wind \in \setint{1}{N}$ as $\bfAt_\wind=\bfF(\rho_\wind)$ and $\bfBt_\wind=\bfG(\rho_\wind)$. 
Worker $\wind\in\setint{1}{N}$ computes the product $\bfAt_\wind \bfBt_\wind$ and returns it to the main node.
In a well-designed scheme, the main node can decode $\bfA_i\bfB_j$, $i\in\setint{1}{K}, j\in\setint{1}{L}$ upon receiving the workers' responses $\bfAt_\wind \bfBt_\wind, \wind\in\setint{1}{N}$.

The task of designing a polynomial code-based PDMM scheme for the OPP then consists in finding vectors of degrees $\ap, \as, \bp, \bs$, and the vector of evaluation points $\boldsymbol{\rho}$, such that the scheme is decodable and $T$-private, while requiring as low a number $N$ of workers as possible.

\subsection{\gaspr and Degree Tables}
The degree table \cite{doliveira2021degreec,doliveira2020GASPa} is a design framework for polynomial codes for PDMM schemes in the OPP. 
Sums between the integers in $\ap, \as$ and $\bp, \bs$ are arranged in an addition table as in \cref{tab:degreetable}. We denote the sets of entries in the top-left, top-right, bottom-left and bottom-right quadrant with $\TL,\TR,\BL,$ and $\BR$, respectively.
\begin{table}[h]
  \centering
  \resizebox{0.72\linewidth}{!}{
\begin{tblr}{vlines, colspec={cc}}
\hline
   $\ap_1\!+\!\bp_1~\dots~\ap_1\!+\!\bp_L$ & $\ap_1\!+\!\bs_1~\dots~\ap_1\!+\!\bs_T$ \\
   $\phantom{\ap_1}~\vdots~\phantom{\bp_1}~~\ddots~\phantom{\ap_1}~~\vdots\phantom{\bp_L}$ & $\phantom{\ap_1}~\vdots~\phantom{\bp_1}~~\ddots~\phantom{\ap_1}~~\vdots\phantom{\bp_L}$ \\
   $\ap_K\!+\!\bp_1~\dots~\ap_K\!+\!\bp_L$ & $\ap_K\!+\!\bs_1~\dots~\ap_K\!+\!\bs_T$ \\ \hline 
   $\as_1\!+\!\bp_1~\dots~\as_1\!+\!\bp_L$ & $\as_1\!+\!\bs_1~\dots~\as_1\!+\!\bs_T$ \\
   $\phantom{\ap_1}~\vdots~\phantom{\bp_1}~~\ddots~\phantom{\ap_1}~~\vdots\phantom{\bp_L}$ & $\phantom{\ap_1}~\vdots~\phantom{\bp_1}~~\ddots~\phantom{\ap_1}~~\vdots\phantom{\bp_L}$ \\
   $\as_T\!+\!\bp_1~\dots~\as_T\!+\!\bp_L$ & $\as_T\!+\!\bs_1~\dots~\as_T\!+\!\bs_T$ \\
\hline
\end{tblr}}
\caption{\small The degree table.}
  \label{tab:degreetable}
\end{table}

Each distinct integer in the degree table corresponds to the degree of a monomial in the polynomial $\bfF(x)\bfG(x)$. Thus, the number $N$ of required workers equals the number of distinct entries in the table.
In~\cite{doliveira2020GASPa, doliveira2021degreec} it is shown that for sufficiently large fields $\bbF_p$ there exists a vector of evaluation points $\boldsymbol{\rho}$ such that the corresponding PDMM scheme is $T$-private and decodable if certain conditions on the degree table hold. 
Any set of up to $T$ workers cannot gain any information about $\bfA$ and $\bfB$ as long as neither $\ap||\as$ nor $\bp||\bs$ contains duplicate entries. 
The main node can decode $\bfA\bfB$ from as many evaluations as there are non-zero coefficients in polynomial $\bfF(x)\bfG(x)$ as long as all $KL$ entries in the top left quadrant of the table are distinct and unique within the table.
We state these requirements formally as follows.
\begin{definition}[Private and Decodable Degree Table \cite{doliveira2020GASPa,doliveira2021degreec}] \label{def:degreetable}
  A tuple $(\ap, \as, \bp, \bs)$ with $\ap \in \bbZ^{K}, \as \in \bbZ^{T}, \bp \in \bbZ^{L}, \bs \in \bbZ^{T}$, is called a private and decodable \emph{degree table} for parameters $K, L$ and $T$ with $N$ distinct entries if the following conditions are fulfilled:\begin{enumerate}[label=\Roman*)]
    \item $|\TL \cup \TR \cup \BL \cup \BR| = N$, 
    \item $|\TL| = K L$,
    \item a) $\TL \cap \TR = \emptyset$, b) $\TL \cap \BL = \emptyset$, c) $\TL \cap \BR = \emptyset$,
    \item $|\{\ap || \as\}| = K+T$, and $|\{\bp||\bs\}| = L+T$.
  \end{enumerate}
  where $\TL = \{\ap\} + \{\bp\}$, $\TR = \{\ap\} + \{\bs\}$, $\BL = \{\as\} + \{\bp\}$, $\BR = \{\as\} + \{\bs\}$.
\end{definition}
We introduce notation for %
generalized arithmetic progressions, which are an important tool in designing degree tables because of their relation to small sumsets, cf.~\cite{tao2006additive}. 
\begin{notation}
  Denote by $\gap(\ell, x, r) \in \bbZ^\ell$ the generalized arithmetic progression of length $\ell$ and chain length $r$, i.e.,
  \begin{align*}
    \gap(\ell, x, r) = (&0, 1, 2, \dots, r-1, \\
    &x, x+1, x+2, \dots, x+r-1,\\ 
                        &2x, 2x+1, 2x + 2, \dots, 2x+r-1, \dots).
  \end{align*}
\end{notation}
When $r=1$, $\gap(\ell, x, 1)=(0, x, 2x, \dots, (\ell-1)x)$ is an arithmetic progression with \emph{common difference} $x$.
Next, we state the construction $\gaspr$ from \cite{doliveira2021degreec}. 
\begin{construction}[\gaspr\cite{doliveira2021degreec}]
  For parameters $K,L,T$ with $K\geq L$, and $1\leq r \leq \min(K,T)$, $\gaspr(K,L,T)$ is defined by $(\ap,\as,\bp,\bs)$ where\begin{equationarray*}{ll}
    \ap = \vecint{0}{K-1},
    &\as = KL + \gap(T, K, r), \\
    \bp = K \cdot \vecint{0}{L-1},
    &\bs = KL + \vecint{0}{T-1}.
  \end{equationarray*}
\end{construction}

The special cases $\gasp_{r=1}$ and $\gasp_{r=\min(K,T)}$ are known as \gaspsmall and \gaspbig, respectively.

\section{Cyclic-Addition Degree Tables} \label{sec:CATframework}

We adapt the degree table by requiring $q$th roots of unity as evaluation points.
This means that the additions in the table are performed over the cyclic group $\bbZ_q$ rather than the integers.

\begin{definition}[Cyclic-Addition Degree Table (CAT)] \label{def:modulodegreetable}
  A tuple $(q, \ap, \as, \bp, \bs)$, where $q$ is a positive integer and $\ap \in \Zq^{K}, \as \in \Zq^{T}, \bp \in \Zq^{L}, \bs \in \Zq^{T}$, is called a \emph{cyclic-addition degree table} for parameters $K, L$ and $T$ with $N$ distinct entries if the following conditions are fulfilled:
  \begin{enumerate}[label=\Roman*)]
    \item $|\TL \cup \TR \cup \BL \cup \BR| = N$, 
    \item $|\TL| = K L$,
    \item a) $\TL \cap \TR = \emptyset$, b) $\TL \cap \BL = \emptyset$, c) $\TL \cap \BR = \emptyset$,
    \item In any prime field $\F_p$ with $q|p-1$, there exist $N$ distinct $q$th roots of unity $\boldsymbol{\rho} = (\rho_1, \dots, \rho_N)$, s.t.\begin{enumerate}
        \item $\bfV(\boldsymbol{\rho}, \boldsymbol{\gamma})$ is invertible and
        \item all $T\times T$ submatrices of $\bfV(\boldsymbol{\rho}, \as)$ and $\bfV(\boldsymbol{\rho}, \bs)$ are invertible,
      \end{enumerate}
  \end{enumerate}
  where $\TL = \{\ap\} + \{\bp\}$, $\TR = \{\ap\} + \{\bs\}$, $\BL = \{\as\} + \{\bp\}$, $\BR = \{\as\} + \{\bs\}$, and $\boldsymbol{\gamma} = \vect(\{\ap || \as\}+\{\bp || \bs\})$, with additions in $\Zq$.
\end{definition}

On a high level, the PDMM scheme corresponding to a \cat is private and decodable for the same reasons as for the conventional degree table. The difference lies in condition IV) which treats the evaluation points explicitly and requires specific matrices to be invertible.
For the conventional degree table, these matrices can be made invertible by picking suitable evaluation points, provided the field $\bbF_p$ is sufficiently large.
Instead of requiring a large field, \cat requires $q|p-1$.

\begin{restatable}{theorem}{thmcat} \label{thm:cat}
  A \cat for parameters $K$, $L$ and $T$ with $N$ distinct entries corresponds to a $T$-private and decodable PDMM scheme for matrices partitioned in $(K, L)$-OPP with $N$ workers. %
\end{restatable}%
The proof of \cref{thm:cat} uses standard techniques and can be found in \ifarxiv \cref{app:thmcat}. \else \cite[\cref{app:thmcat}]{arxivversion}. \fi
\Cref{fig:GASP444_v_CAT444} shows a \cat next to a \gaspr degree table. 

\begin{figure}[t]
\centering
\begin{subfigure}{0.48\linewidth}
    \centering
    \resizebox{.75\textwidth}{!}{
    \input{tikz/CAT_444.tex}}
    \caption{\small CAT$_{x=3}$ with $N=34$ distinct entries.}
    \label{fig:CAT444}
\end{subfigure}
\hfill
\begin{subfigure}{0.48\linewidth}
    \centering
    \resizebox{.75\textwidth}{!}{
    \input{tikz/GASP_444.tex}}
    \caption{\small GASP$_{r=2}$ with $N=36$ distinct entries.}
    \label{fig:GASP444}
\end{subfigure}
\caption{\small A CAT and the best \gaspr degree table for $K=L=T=4$. The first column corresponds to $\ap || \as$; the first row to $\bp || \bs$. The other entries are the (modulo-$q$/integer) sums of the corresponding entries in the first row and column.}
\label{fig:GASP444_v_CAT444}
\vspace{-1em}
\end{figure}

\begin{example}  \label{ex:continued_introexample}
  We continue the example from \cref{sec:introexample} for $K=L=T=2$ with $q=10$, $\ap= (0, 3)$, $\as=(6, 7)$, $\bp=(0, 1)$ and $\bs=(9, 2)$.
  In \cref{fig:introcat222} we can see that $\TL=\{0, 1, 3, 4\}$, $\TR=\{2, 5, 9\}$, $\BL=\{6, 7, 8\}$ and $\BR=\{5, 6, 8, 9\}$.
  Condition I) guarantees that $\bfH(x)$ has $N$ non-zero coefficients, and thus that $\bfV(\boldsymbol{\rho}, \boldsymbol{\gamma})$ is square. The set $\{\boldsymbol{\gamma}\} = \TL \cup \TR \cup \BL \cup \BR$ is the set of integers in the table; in the example $\setint{0}{9}$.
  Condition II) guarantees that the desired products $\bfA_1 \bfB_1$,  $\bfA_1 \bfB_2$,  $\bfA_2 \bfB_1$,  $\bfA_2 \bfB_2$ occur in distinct coefficients in $\bfH(x)$, i.e., they are not mixed with each other in the polynomial multiplication. Condition III) makes sure these desired products are not mixed with random matrices either.
  Condition IV) a) guarantees, that the coefficients of $\bfH(x)$ can be recovered from the evaluations at $\boldsymbol{\rho}$. In the example we have
  \ifarxiv \vspace{-0.5em} \fi
    \begin{align*}
  \bfV(\boldsymbol{\rho}, \boldsymbol{\gamma}) &= \left(\begin{smallmatrix}
    \omega^0 & \omega^0 & \omega^0 & \dots & \omega^0  \\
    \omega^0 & \omega^1 & \omega^2 & \dots & \omega^9  \\
    \omega^0 & \omega^2 & \omega^4 & \dots & \omega^8  \\
    \omega^0 & \omega^3 & \omega^6 & \dots & \omega^7  \\
    \vdots & \vdots & \vdots & \ddots & \vdots  \\
    \omega^0 & \omega^9 & \omega^8 & \dots & \omega^1  \\
  \end{smallmatrix}\right) %
   = \left(\begin{smallmatrix}
     1 & 1 & 1& \dots & 1 \\
     1 & 2 & 4& \dots & 6 \\
     1 & 4 & 5& \dots & 3 \\
     1 & 8 & 9& \dots & 7 \\
    \vdots & \vdots & \vdots & \ddots & \vdots  \\
     1 & 6 & 3& \dots & 2 \\
  \end{smallmatrix}\right),
\end{align*}
which is invertible due to its Vandermonde structure. 

Note that even if the exponents of non-zero coefficients in $\bfH(x)$ are not consecutive, if $\boldsymbol\rho$ consists of consecutive powers of a primitive $q$th root of unity, $\bfV(\boldsymbol{\rho}, \boldsymbol{\gamma})$ is still invertible since then it is the transpose of a Vandermonde matrix.

Condition IV) b) guarantees information theoretic $T$-privacy of $\bfA$ and $\bfB$ by ensuring that the linear combinations of random matrices used to obscure the blocks of $\bfA$ and $\bfB$ 
are linearly independent for any $T$ workers' tasks. As a consequence, they are also statistically independent and uniformly distributed.
In the example we have 
{%
  \begin{align*}\bfV(\boldsymbol{\rho}, \as) &= \left(\begin{smallmatrix} 
  \omega^0 & \omega^0 \\
  \omega^6 & \omega^7 \\
  \omega^2 & \omega^4 \\
  \vdots & \vdots \\
  \omega^4 & \omega^3
  \end{smallmatrix}\right) %
  = \left(\begin{smallmatrix} 
   1 &  1\\
   9 &  7\\
   4 &  5\\
   \vdots & \vdots \\
   5 &  8
 \end{smallmatrix}\right).
\end{align*}
}%
As shown in \cref{sec:introexample}, any $2\times 2$ submatrix is invertible.
The same holds for $\bfV(\boldsymbol{\rho}, \bs)$.

\end{example}

\section{A Family of CATs} %
\label{sec:family}
Throughout this section let $K\geq L \geq T \geq 2$.
We provide an explicit construction of \cats after introducing some notation.
\begin{definition} \label{def:KLTshorthand}
  For integers $K \geq L \geq T \geq 2$, let $\kappa$ and $\lambda$ be the smallest non-negative integers, such that $K+1+\kappa$ and $L+1+\lambda$ are co-prime to $T-1$.
  We further define the following shorthand notation 
  \ifarxiv \vspace{-0.5em} \fi
  \begin{align*}
    \Ks \defeq K+1+\kappa,\quad \Ls \defeq L+1+\lambda, \quad \Tm \defeq T-1.
  \end{align*}
  \ifarxiv \vspace{-0.5em} \fi
\end{definition}

\begin{construction}[\catx] \label{constr:cat}
  For integers $K \geq L \geq T \geq 2$, let 
  \ifarxiv \vspace{-0.5em} \fi
  \begin{align*}
  q = \Ks \Ls + \Tm^2.
\end{align*}
  Let $x$ be a positive integer, s.t. $x\perp q$ and let $y$ be the solution\footnote{Existence and uniqueness follow from \cref{claim:merged} in the sequel.} to $x\Tm +y\Ks \equiv 0 \pmod q$.
  $\catx(K, L, T)$ is defined by $(q, \ap, \as, \bp, \bs)$, where%
  \begin{align*}
    \ap &= y \cdot \vecint{0}{K-1} \bmod q, \\
    \as &= x \cdot \vecint{0}{T-1} + Ky \bmod q, \\
    \bp &= x \cdot \vecint{0}{L-1} \bmod q, \\
    \bs &= y \cdot \vecint{0}{T-1} - x \bmod q.
  \end{align*}
\end{construction}
\begin{remark}
Many suitable choices for $x$ exist, e.g., $x=1$, $x=\Ks$, $x=\Ls$, and $x=\Tm$.
They all lead to valid \cats with the same number of workers, as shown in the sequel.
\end{remark}
\Cref{fig:CAT444} shows $\cat_{x=3}(4, 4, 4)$. Equivalent CATs for different values of $x$ can be found in \cref{fig:CATxs444} in 
\ifarxiv \cref{app:supplementaryfigures}. \else \cite[\cref{app:supplementaryfigures}]{arxivversion}. \fi

Before stating that \catx is indeed a valid \cat, we provide a result relating to the choice of $x$, which is proven in \ifarxiv \cref{app:proofclaimmerged}. \else \cite[\cref{app:proofclaimmerged}]{arxivversion}. \fi

\begin{restatable}{claim}{claimmerged} \label{claim:merged}
  For a given $q$, as in \cref{constr:cat}, of the form $q = \Ks\Ls+ \Tm^2$ with $\Ks$ and $\Ls$ coprime to $\Tm$, 
  it holds that $\Ks \perp q$, $\Ls \perp q$ and $\Tm \perp q$. 
  For any integer $x\perp q$ there exists a unique integer $y\in\setint{0}{q-1}$ s.t. $x\Tm +y\Ks \equiv 0 \pmod q$. Further it holds that $y \perp q$.%
\end{restatable}

\ifarxiv \vspace{-0.5em} \fi
\begin{restatable}{theorem}{thmcatx} \label{thm:catx}
  \Cref{constr:cat} is a CAT with number of distinct entries $\ncatx(K, L, T) = (K+1)(L+1)+(T-1)^2 + \kappa + \lambda$, with $\kappa$ and $\lambda$ as in \cref{def:KLTshorthand}.
\end{restatable}
\ifarxiv \pagebreak \fi

The proof of \cref{thm:catx} can be found in \ifarxiv \cref{app:thmcatx} \else \cite[\cref{app:thmcatx}]{arxivversion} \fi and depends on the following lemma which is proven in \ifarxiv \cref{app:proofarithmeticprogressions}. \else \cite[\cref{app:proofarithmeticprogressions}]{arxivversion}. \fi

\begin{restatable}{lemma}{lemarithmeticprogressions} \label{lemma:arithmeticprogressions}
  Condition IV) in \cref{def:modulodegreetable} is fulfilled if $\as$ and $\bs$ are arithmetic progressions with common differences coprime to $q$.
\end{restatable}

An analogous argument points to an explicit way of selecting evaluation points for \gaspsmall and \gaspbig.

\begin{corollary}
  When constructed over a field $\bbF_p$ with $\omega^0, \omega^1, \dots, \omega^{N-1}$ as evaluation points, \gaspsmall is decodable and private if a) $q$ is larger than the largest entry in the degree table; b) $\omega$ is a primitive $q$th root of unity in $\bbF_p$ (implying $q|p-1$); and c) $q\perp K$.
  For \gaspbig conditions a) and b) are sufficient.
\end{corollary}

\catx saves $3T-5$ workers over \gaspsmall whenever $K+1\perp T-1$, and $L+1\perp T-1$, as can be seen by comparing \cref{thm:catx} to \cite[Table~I]{doliveira2020GASPa}. In general, the difference is $\ngaspsmall(K,L,T) - \ncatx(K,L,T) = 3T-5-\kappa-\lambda$.
The numbers $\kappa$ and $\lambda$ are typically small as illustrated by the following observation. The probability that two random integers less than $n$ are coprime tends to $6/\pi^2 > 60\%$ as $n \to \infty$~\cite[Theorem~332]{hardy1960introduction}. The values of $\kappa$ (and $\lambda$) for various $K$ and $T$ ($L$ and $T$) can be seen in \cref{fig:kappalambda} in \ifarxiv \cref{app:supplementaryfigures}. \else \cite[\cref{app:supplementaryfigures}]{arxivversion}. \fi

\section{New Constructions of Degree Tables} \label{sec:nonmod}

We introduce two new constructions of degree tables. 
Both use two parameters, $r$, and $s$, that must be chosen appropriately. In both cases, as in \gaspr, no general closed-form expression is known for the parameter values minimizing $N$. In our numerical results they are found by computer search.

We generalize \gaspr by introducing a second chain length parameter $s$ and replacing $\bs$ by a generalized arithmetic progression. When $s=T$ the construction equals $\gaspr$.
\begin{construction}[\gasprs] \label{constr:gasprs}
  For integers $K,L,T \geq  2$ and $r,s \leq \min(K,T)$ let $\gasprs(K, L, T)$ be defined by
  \begin{equationarray*}{ll}
    \ap = \vecint{0}{K-1},         &\as = KL + \gap(T, K, r), \\
    \bp = \vecint{0}{L-1} \cdot K, &\bs = KL + \gap(T, K, s).
  \end{equationarray*}
\end{construction}%
 
The following construction is similar to \gasprs, in that $\ap$ and $\bp$ are arithmetic progressions, and $\as$ and $\bs$ are generalized arithmetic progressions.
The specific values, however, were found through a combination of manual and computer search. We call this construction \textbf{d}iscretely \textbf{o}ptimized \textbf{G}ASP$_\textsubscript{rs}$ or \dogrs for short.\footnote{We thank an anonymous ISIT reviewer for pointing out that DOG$_{r=T,s=T}(K,L,T)$ equals the $A3S$ scheme from~\cite{kakar2019capacity}.} 

\begin{construction}[\dogrs] \label{constr:nomodxy}
  For integers, $K, L, T \geq 2$, $1\leq r \leq T$, and $1\leq s \leq \min(T, K+r)$, let $\dogrs(K, L, T)$ be defined by
  \begin{equationarray*}{ll}
    \ap = \vecint{0}{K-1},             &\as = K + \gap(T, K+r, r), \\
    \bp = (K+r)\cdot\vecint{0}{L-1}, &\bs = (K+r)(L-1)+K \\ &\quad\quad\quad\quad+\gap(T, K+r, s).
  \end{equationarray*}
\end{construction}

Let $\ngasprs(K,L,T)$ and $\ndogrs(K,L,T)$ equal the number of distinct entries in $\gasprs(K,L,T)$ and $\dogrs(K,L,T)$, respectively.
While closed-form expressions for $\ngasprs(K,L,T)$ and $\ndogrs(K,L,T)$ can be derived using the inclusion-exclusion principle, (cf. \cite[Sections V.B., VI.B.]{doliveira2020GASPa}) they are not shown in this paper since the expressions are highly discontinuous and of considerable complexity.
The following lemmas are proven in \ifarxiv \cref{app:lemmagasprsvalid} and \cref{app:lemmadogrsvalid}. \else \cite[\cref{app:lemmagasprsvalid}]{arxivversion} and \cite[\cref{app:lemmadogrsvalid}]{arxivversion}. \fi %

\begin{restatable}{lemma}{lemmagasprsvalid} \label{lemma:gasprsvalid}
  $\gasprs(K, L, T)$ is a $T$-private and decodable degree table with $\ngasprs(K,L,T)$ distinct entries satisfying \cref{def:degreetable}.
\end{restatable}
\vspace{-0.5em}
\begin{restatable}{lemma}{lemmadogrsvalid} \label{lemma:dogrsvalid}
  $\dogrs(K, L, T)$ is a $T$-private and decodable degree table with $\ndogrs(K,L,T)$ distinct entries satisfying \cref{def:degreetable}.
\end{restatable}
\vspace{-0.5em}

\section{Numerical Comparison}

We compare the numbers of workers used by \catx, \gasprs and \dogrs to the state-of-the-art, \polegap and \gaspr. %

\Cref{fig:bestschemeKisL} shows the scheme using fewest workers for $2 \leq K=L \leq 20$ and $2 \leq T \leq 20$. Results for $K\neq L$ can be found in \cref{fig:bestschemes_KneqL} in \ifarxiv \cref{app:supplementaryfigures}. \else  \cite[\cref{app:supplementaryfigures}]{arxivversion}. \fi
\gasprs, while it does outperform every other scheme (including \gaspr) \emph{individually} for some parameters, has not been observed to outperform all of them \emph{simultaneously}. 
The best $s$ for \gasprs is shown in \cref{fig:GASPrs_best_s_heatmap} in \ifarxiv \cref{app:supplementaryfigures}. \else \cite[\cref{app:supplementaryfigures}]{arxivversion}.  \fi

Heuristically, we observe that whenever $T$ is much smaller than $K$ and $L$, \catx uses the lowest number of workers. As $T$ increases, \dogrs is dominant.  This regime includes $K=L=T$, when they are large. As $T$ further grows, \polegap uses the least number of workers when it is defined (i.e., $K$ or $L$ is even); otherwise, and when $T$ grows very large, \gaspr is the scheme using the fewest workers.
\vspace{-0.5em}

\begin{figure}[tbh]
  \resizebox{\linewidth}{!}{\input{tikz/KLTplot.tex}}
  \caption{\small The relative improvement in number of workers of \dogrs and \gasprs over \gaspr for $2\leq K=L=T \leq 100$. One point where $\ndogrs(3, 3, 3)=23$ and $\ngaspr(3, 3,3 )=22$ (i.e. $-4.3\%$) is outside the visible area.}
    \label{fig:relimprovement}
\end{figure}

Figure~\ref{fig:relimprovement} shows the relative number of workers saved by \dogrs and \gasprs over \gaspr. We observe that for $30 \leq K=L=T \leq 100$, both schemes save around $5\%$. %

\section{Conclusion} \label{sec:conclusion}

We have demonstrated improved performance in PDMM, using roots of unity in conjunction with the degree table. Additionally, we have presented new and improved constructions of degree tables without using roots of unity, demonstrating in a second way that there is still room for improvement in PDMM for the OPP.
The derivation of converse results as well as a detailed asymptotic analysis are left for future work.

\clearpage
\bibliographystyle{IEEEtran}
\bibliography{references,ref_arxiv}

\clearpage
\appendix

\subsection{Proof of \cref{thm:cat}} \label{app:thmcat}

\thmcat*
\begin{proof}[Proof of \cref{thm:cat}]
  We prove privacy and decodability using standard techniques in PDMM, cf. \cite{chang2018capacity}.

  \emph{Privacy:} 
  Let $\cT\subset \setint{1}{N}$ denote any set of $|\cT|=T$ worker nodes and let $\rvAtT\defeq \{\bfF(\rho_t) | t \in \cT\}$ and $\rvBtT\defeq \{\bfG(\rho_t) | t \in \cT\}$ denote the corresponding tasks sent to these workers. We use $\bfR_{\setint{1}{T}}$ as shorthand notation for $\bfR_1, \dots, \bfR_T$.

  Let $\bfA$ and $\bfB$ be distributed according to some unknown distribution.
  We point out the following key properties of the scheme that enable the proof: 1) the encoding of the tasks is deterministic given $\bfA$ and the random matrices $\bfR_{\setint{1}{T}}$, i.e., $H(\bfAtT| \bfA, \bfR_{\setint{1}{T}}) = 0$; and 2) from condition IV) b) and the structure of the encoding it follows that given $\bfA$ and $\rvAtT$, the random matrices $\bfR_{\setint{1}{T}}$ can be recovered by solving a linear system, i.e., $H(\bfR_{\setint{1}{T}} | \bfA, \rvAtT)=0$.
\\
  We have\begin{align}
    \En(\rvAtT | \rvA) &= \MI(\bfR_{\setint{1}{T}}; \rvAtT | \rvA) + \En(\rvAtT | \rvA, \bfR_{\setint{1}{T}}) \label{eq:thmcat1}  \\ 
                       &= \MI(\bfR_{\setint{1}{T}}; \rvAtT | \rvA)  \label{eq:thmcat2}  \\
                       &= \En(\bfR_{\setint{1}{T}} | \rvA) - \En(\bfR_{\setint{1}{T}} | \rvA, \rvAtT)  \label{eq:thmcat3}  \\
                       &= \En(\bfR_{\setint{1}{T}} | \rvA)  \label{eq:thmcat4}  \\
                       &= \En(\bfR_{\setint{1}{T}})  \label{eq:thmcat5}  \\
                       &\geq \En(\bfAtT),  \label{eq:thmcat6}  
  \end{align}
  where \eqref{eq:thmcat1} and \eqref{eq:thmcat3} follow from the definition of mutual information; \eqref{eq:thmcat2} and $\eqref{eq:thmcat4}$ follow from $H(\bfAtT| \bfA, \bfR_{\setint{1}{T}}) = 0$ and $H(\bfR_{\setint{1}{T}} | \bfA, \rvAtT)=0$, respectively; and \eqref{eq:thmcat5} and \eqref{eq:thmcat6} follow because $\bfR_{\setint{1}{T}}$ is independent of $\rvA$ and maximum entropy.

  Since mutual information is non-negative, it holds that $\MI(\rvAtT; \rvA) = \En(\rvAtT) - \En(\rvAtT | \rvA) = 0$.
  $\MI(\rvB; \rvBtT) = 0$ follows from the analogous arguments. 

\emph{Decodability: } Decodability follows from IV) a). Given $N$ evaluations the main node can interpolate $\bfH(x) \defeq \bfF(x)\bfG(x) \pmod{x^q-1}$ by inverting $\bfV(\boldsymbol{\rho}, \boldsymbol{\gamma})$ to recover the coefficients of $\bfH(x)$. By conditions II) and III), there are $KL$ coefficients that equal exactly $\bfA_i \bfB_j$ for all $1\leq i \leq K, 1\leq j \leq L$.
\end{proof}

\subsection{Proof of \cref{claim:merged}} \label{app:proofclaimmerged}
\claimmerged*
\begin{proof}
  \emph{$\Ks,\Ls,$ and $\Tm$ are coprime to $q$:}
  Assume $\Ks$ and $q$ have a common prime factor $u$. Then, the number 
  \begin{align*}
    \frac{\Ks\Ls + \Tm^2}{u} =
    \frac{\Ks}{u}\Ls + \frac{\Tm^2}{u}
  \end{align*}
    is an integer, implying %
    $u \mid \Tm$. A contradiction.
  The statements $\Ls \perp q$ and $\Tm \perp q$ follow from analogous arguments.

  \emph{Existence and uniqueness of $y$:}
  Since $\Ks \perp q$, its inverse $(\Ks)^{-1} \bmod q$ in $\Zq$ exists and is unique. Then, $y = -x\Tm (\Ks)^{-1} \bmod q$.

  \emph{$y\perp q$:}
  Assume there exists a prime $u$, that divides both $q$ and $y$.
  From $x\Tm + y\Ks \equiv 0 \pmod q$, there exists an integer $a$, s.t. $aq = x\Tm + y\Ks$, i.e.,
  \begin{align*}
    a\frac{q}{u} - \frac{y\Ks}{u} &= \frac{x\Tm}{u}.
  \end{align*}
  The left hand side is clearly integer, thus $u$ must divide $x$ or $\Tm$, contradicting $x\perp q$ or $\Tm \perp q$.
\end{proof}

\subsection{Proof of \cref{thm:catx}} \label{app:thmcatx}

\thmcatx*
In the following, we state four lemmas that contribute to the proof of \cref{thm:catx}, which is presented thereafter. The proofs of the lemmas can be found in separate appendices for improved clarity. %

The proof contains some geometric elements which are illustrated in \cref{fig:proofintervalslattice}.

\begin{figure}
  \resizebox{\linewidth}{!}{\begin{tikzpicture}%
\begin{axis}[point meta max={nan}, point meta min={nan}, legend cell align={left}, legend columns={1}, 
  legend style={font={{\fontsize{8 pt}{10.4 pt}\selectfont}}, 
  at={(0.98, 0.5)}, anchor={east}}, 
  axis background/.style={fill={rgb,1:red,1.0;green,1.0;blue,1.0}, opacity={1.0}}, anchor={north west}, xshift={1.0mm}, yshift={-1.0mm}, 
  width={90mm}, height={60mm}, 
  scaled x ticks={false}, xlabel={$i$}, x tick style={color={rgb,1:red,0.0;green,0.0;blue,0.0}, opacity={1.0}}, x tick label style={color={rgb,1:red,0.0;green,0.0;blue,0.0}, opacity={1.0}, rotate={0}}, xlabel style={at={(ticklabel cs:0.5)}, anchor=near ticklabel, at={{(ticklabel cs:0.5)}}, anchor={near ticklabel}, font={{\fontsize{11 pt}{14.3 pt}\selectfont}}, color={rgb,1:red,0.0;green,0.0;blue,0.0}, draw opacity={1.0}, rotate={0.0}}, xmajorgrids={false}, xmin={-7}, xmax={16}, xticklabels={{$-10$,$-5$,$0$,$5$,$10$,$15$}}, xtick={{-10.0,-5.0,0.0,5.0,10.0,15.0}}, xtick align={inside}, xticklabel style={font={{\fontsize{8 pt}{10.4 pt}\selectfont}}, color={rgb,1:red,0.0;green,0.0;blue,0.0}, draw opacity={1.0}, rotate={0.0}}, x grid style={color={rgb,1:red,0.0;green,0.0;blue,0.0}, draw opacity={0.1}, line width={0.5}, solid}, axis x line*={left}, x axis line style={color={rgb,1:red,0.0;green,0.0;blue,0.0}, draw opacity={1.0}, line width={1}, solid}, scaled y ticks={false}, ylabel={$j$}, y tick style={color={rgb,1:red,0.0;green,0.0;blue,0.0}, opacity={1.0}}, y tick label style={color={rgb,1:red,0.0;green,0.0;blue,0.0}, opacity={1.0}, rotate={0}}, ylabel style={at={(ticklabel cs:0.5)}, anchor=near ticklabel, at={{(ticklabel cs:0.5)}}, anchor={near ticklabel}, font={{\fontsize{11 pt}{14.3 pt}\selectfont}}, color={rgb,1:red,0.0;green,0.0;blue,0.0}, draw opacity={1.0}, rotate={0.0}}, ymajorgrids={false}, ymin={-10}, ymax={10}, yticklabels={{$-10$,$-5$,$0$,$5$,$10$}}, ytick={{-10.0,-5.0,0.0,5.0,10.0}}, ytick align={inside}, yticklabel style={font={{\fontsize{8 pt}{10.4 pt}\selectfont}}, color={rgb,1:red,0.0;green,0.0;blue,0.0}, draw opacity={1.0}, rotate={0.0}}, y grid style={color={rgb,1:red,0.0;green,0.0;blue,0.0}, draw opacity={0.1}, line width={0.5}, solid}, axis y line*={left}, y axis line style={color={rgb,1:red,0.0;green,0.0;blue,0.0}, draw opacity={1.0}, line width={1}, solid}, colorbar={false}]
    \addplot[color={rgb,1:red,0.0;green,0.0;blue,0.0}, name path={19}, only marks, draw opacity={1.0}, line width={0}, solid, mark={*}, mark size={2pt}, mark repeat={1}, mark options={color={rgb,1:red,0.0;green,0.0;blue,0.0}, draw opacity={1.0}, fill={rgb,1:red,0.0;green,0.0;blue,0.0}, fill opacity={1.0}, line width={0.75}, rotate={0}, solid}]
        table[row sep={\\}]
        {
            \\
            -14.0  -22.0  \\
            20.0  -22.0  \\
            -1.0  -21.0  \\
            -22.0  -20.0  \\
            12.0  -20.0  \\
            -9.0  -19.0  \\
            4.0  -18.0  \\
            -17.0  -17.0  \\
            17.0  -17.0  \\
            -4.0  -16.0  \\
            9.0  -15.0  \\
            -12.0  -14.0  \\
            22.0  -14.0  \\
            1.0  -13.0  \\
            -20.0  -12.0  \\
            14.0  -12.0  \\
            -7.0  -11.0  \\
            6.0  -10.0  \\
            -15.0  -9.0  \\
            19.0  -9.0  \\
            -2.0  -8.0  \\
            11.0  -7.0  \\
            -10.0  -6.0  \\
            3.0  -5.0  \\
            -18.0  -4.0  \\
            16.0  -4.0  \\
            -5.0  -3.0  \\
            8.0  -2.0  \\
            -13.0  -1.0  \\
            21.0  -1.0  \\
            0.0  0.0  \\
            -21.0  1.0  \\
            13.0  1.0  \\
            -8.0  2.0  \\
            -16.0  4.0  \\
            18.0  4.0  \\
            10.0  6.0  \\
            -11.0  7.0  \\
            2.0  8.0  \\
            -19.0  9.0  \\
            15.0  9.0  \\
            -6.0  10.0  \\
            7.0  11.0  \\
            -14.0  12.0  \\
            20.0  12.0  \\
            -1.0  13.0  \\
            -22.0  14.0  \\
            12.0  14.0  \\
            -9.0  15.0  \\
            4.0  16.0  \\
            -17.0  17.0  \\
            17.0  17.0  \\
            -4.0  18.0  \\
            9.0  19.0  \\
            -12.0  20.0  \\
            22.0  20.0  \\
            1.0  21.0  \\
            -20.0  22.0  \\
            14.0  22.0  \\
        }
        ;
    \addlegendentry {$ix\equiv jy$}
    \addplot[color={rgb,1:red,0.0;green,0.0;blue,1.0}, name path={21}, only marks, draw opacity={1.0}, line width={0}, solid, mark={*}, mark size={2pt}, mark repeat={1}, mark options={color={rgb,1:red,0.0;green,0.0;blue,0.0}, draw opacity={1.0}, fill={rgb,1:red,1.0;green,1.0;blue,1}, fill opacity={1.0}, line width={0.0}, rotate={0}, solid}]
        table[row sep={\\}]
        {
            \\
            0.0  0.0  \\
        }
        ;
    \addlegendentry {0}
    \addplot[color={rgb,1:red,1.0;green,0.0;blue,0.0}, name path={22}, only marks, draw opacity={1.0}, line width={0}, solid, mark={square*}, mark size={2pt}, mark repeat={1}, mark options={color={rgb,1:red,0.0;green,0.0;blue,0.0}, draw opacity={1.0}, 
   fill={rgb,1:red,0.0667;green,0.4392;blue,0.6667},
  fill opacity={1.0}, line width={0.0}, rotate={0}, solid}]
        table[row sep={\\}]
        {
            \\
            -3.0  5.0  \\
        }
        ;
    \addlegendentry {\sol1}
    \addplot[color={rgb,1:red,1.0;green,0.7529;blue,0.7961}, name path={23}, only marks, draw opacity={1.0}, line width={0}, solid, mark={diamond*}, mark size={3pt}, mark repeat={1}, mark options={color={rgb,1:red,0.0;green,0.0;blue,0.0}, draw opacity={1.0}, 
  fill={rgb,1:red,0.9882;green,0.4902;blue,0.0431},   
  fill opacity={1.0}, line width={0.0}, rotate={0}, solid}]
        table[row sep={\\}]
        {
            \\
            5.0  3.0  \\
        }
        ;
    \addlegendentry {\sol2}
    \addplot[color={rgb,1:red,0.1216;green,0.4667;blue,0.7059}, name path={24}, area legend, fill={none}, fill opacity={1}, draw opacity={1.0}, line width={1}, solid, on layer=foreground]
        table[row sep={\\}]
        {
            \\
            -3.0  -3.0  \\
            -1.0  -3.0  \\
            -1.0  3.0  \\
            -3.0  3.0  \\
            -3.0  -3.0  \\
        }
        ;
    \addlegendentry {II)}
    \addplot[color={rgb,1:red,1.0;green,0.498;blue,0.0549}, name path={25}, area legend, fill={none}, fill opacity={1}, draw opacity={1.0}, line width={1}, dashdotted, on layer=foreground]
        table[row sep={\\}]
        {
            \\
            -4.0  -6.0  \\
            -1.0  -6.0  \\
            -1.0  3.0  \\
            -4.0  3.0  \\
            -4.0  -6.0  \\
        }
        ;
    \addlegendentry {III) a}
    \addplot[color={rgb,1:red,0.1725;green,0.6275;blue,0.1725}, name path={26}, area legend, fill={none}, fill opacity={1}, draw opacity={1.0}, line width={1}, dashed, on layer=foreground]
        table[row sep={\\}]
        {
            \\
            -3.0  -4.0  \\
            6.0  -4.0  \\
            6.0  -1.0  \\
            -3.0  -1.0  \\
            -3.0  -4.0  \\
        }
        ;
    \addlegendentry {III) b}
    \addplot[color={rgb,1:red,0.8392;green,0.1529;blue,0.1569}, name path={27}, area legend, fill={none}, fill opacity={1}, draw opacity={1.0}, line width={1}, dotted,on layer=foreground]
        table[row sep={\\}]
        {
            \\
            -4.0  -7.0  \\
            2.0  -7.0  \\
            2.0  -1.0  \\
            -4.0  -1.0  \\
            -4.0  -7.0  \\
        }
        ;
    \addlegendentry {III) c}
    \addplot[color={rgb,1:red,0.0;green,0.0;blue,0.0}, name path={20}, area legend, fill={rgb,1:red,0.76;green,0.76;blue,0.76}, fill opacity={1}, draw opacity={1}, line width={0}, solid,on layer=main]
        table[row sep={\\}]
        {
            \\
            -4.0  -7.0  \\
            7.0  -7.0  \\
            7.0  3.0  \\
            -4.0  3.0  \\
            -4.0  -7.0  \\
        }
        ;
      \addlegendentry {\cref{lemma:boundingbox}}
\end{axis}
\end{tikzpicture}}
  \caption{\small Visualization of the various intervals as well as solutions to $ix\equiv jy\pmod q$ involved in the proof of \cref{thm:cat}. For the example of $K=L=T=4$.}
  \label{fig:proofintervalslattice}
\end{figure}

\begin{restatable}{lemma}{lemmaequivconstraints} \label{lemma:equivconstraints}
  \cref{constr:cat} fulfills conditions II) and III) of \cref{def:modulodegreetable} if $ix \not\equiv jy \pmod q$
      for all pairs $i, j$ given by \begin{itemize}
        \item $i \in \setint{-L}{-1}, j\in \setint{-K-T+2}{K-1}$, 
        \item $i \in \setint{-L+1}{T+L-2}, j\in \setint{-K}{-1}$,
        \item $i \in \setint{-L}{T-2}, j\in \setint{-K-T+1}{-1}$.
      \end{itemize}
\end{restatable}

\begin{restatable}{lemma}{lemmalatticesolutions} \label{lemma:latticesolutions}
  Any solution to the equation $ix \equiv jy \pmod q$ can be expressed as
  $(i = -a\Tm+b\Ls, j = a\Ks+b\Tm)$,
  for some integers $a$ and $b$.
\end{restatable}

Note that the set $\{(i, j) | ix+jy \equiv  0 \pmod q\}$ is the same for all $x\perp q$ and accompanying $y$.

\begin{restatable}{lemma}{lemmaboundingbox} \label{lemma:boundingbox}
  No solutions to $ix \equiv jy \pmod q$ fulfill $-L \leq i \leq T+L-1$ and $-K-T+1 \leq j \leq K-1$, other than $(i=0,j=0), (i=\Ls, j=\Tm),$ and $(i=\Tm, j=-\Ks)$.
\end{restatable}
Note that the third point $(i=\Tm, j=-\Ks)$ can be inside or outside these intervals depending on the parameters $K, L$, and $T$.

\begin{restatable}{lemma}{lemmaintersections} \label{lemma:intersections}
  In \catx, as defined in \cref{constr:cat}, it holds that\begin{align*}
    |\TR \cap \BR| = 2\Tm - \kappa \text{ and } 
    |\BL \cap \BR| = 2\Tm - \lambda.
  \end{align*}
\end{restatable}

\begin{proof}[Proof of \cref{thm:catx}]
  We show that the conditions in \cref{def:modulodegreetable} are fulfilled one-by-one.

  \emph{I):}
    First, we count the number of distinct integers in each quadrant separately. 
    We have\begin{equationarray*}{ll}
      |\TL| = KL,    &|\TR| = K+T-1,\\ 
      |\BL| = L+T-1, &|\BR| = T^2, 
    \end{equationarray*} where $\TR$ and $\TL$ are sumsets of arithmetic progressions with the same common difference (c.f. \cite[Proposition~5.8, Theorem~5.9]{tao2006additive}) and $\BR$ is a translated subset of $\TL$.
    
    Next, we establish that $\TR \cap \BL = \emptyset$. Assume not, then there exist  integers $1\leq i \leq L+T-1$ and $-K \leq j \leq T-2$, s.t. $jy \equiv ix \pmod q$, contradicting \cref{lemma:boundingbox}.

Thus, $\ncatx(K,L,T) = |\TL| + |\TR| + |\BL| + |\BR| - |\TR \cap \BR| - |\BL \cap \BR|$ by the inclusion-exclusion principle.
  From \cref{lemma:intersections} we know that $|\TR \cap \BR| = 2\Tm - \kappa$ and $|\BL \cap \BR| = 2\Tm - \lambda$.
  After simplifying we get\begin{align*}
    \ncatx(K,L,T) %
                  &= (K+1)(L+1) + (T-1)^2 + \kappa + \lambda.
  \end{align*}
  \emph{II) and III):} 
  It is readily verified, that the intervals given in \cref{lemma:boundingbox} contain the ones in \cref{lemma:equivconstraints} and further that the solutions to $ix \equiv jy \pmod q$ given in \cref{lemma:boundingbox} do not intersect the intervals stated in \cref{lemma:equivconstraints}.

    \emph{IV):} The condition is fulfilled by \cref{lemma:arithmeticprogressions} since \as and \bs are arithmetic progressions with common difference $x$ and $y$, respectively, and $x,y$ are coprime to $q$ (cf. \cref{claim:merged}).
\end{proof}

\subsection{Proof of \cref{lemma:arithmeticprogressions}} \label{app:proofarithmeticprogressions}

\lemarithmeticprogressions*
\begin{proof}
  Let $g\in \bbF_p$ be a generator of $\bbF_p$ and let $\omega=g^{\frac{p-1}{q}}$. Then, the order of $\omega$ in $\bbF_p$ is $q$, i.e. $\omega^i \neq 1$ for $i \in \setint{1}{q-1}$ and $\omega^q=1$.
  Let $\boldsymbol{\rho} = (\omega^0, \omega^1, \dots, \omega^{N-1})$. 
  \\
  \emph{IV) a):} The matrix
  \begin{align*} 
    \bfV(\boldsymbol{\rho}, \boldsymbol{\gamma}) &= \left(\omega^{i \gamma_j}\right)_{0\leq i \leq N-1, 1 \leq j \leq N} \\
                                                 &= \left((\omega^{\gamma_j})^i\right)_{0\leq i \leq N-1, 1 \leq j \leq N}
  \end{align*}
  is a transposed Vandermonde matrix and $\omega^{\gamma_1}, \omega^{\gamma_2}, \dots, \omega^{\gamma_N}$ are distinct since $\omega$ is of order $q$ and $\gamma_1, \gamma_2, \dots, \gamma_N$ are distinct integers from $\setint{0}{q-1}$.
  \\
  \emph{IV) b):} 
  Let $\asind{j} = b_\alpha + x_\alpha j$.
  It holds that
  \begin{align*} 
    \bfV(\boldsymbol{\rho}, \as) &= (\omega^{i \asind{j}})_{0\leq i \leq N-1, 1 \leq j \leq T} \\
                                 &= (\omega^{i (x_\alpha j + b_\alpha)})_{0\leq i \leq N-1, 1 \leq j \leq T} \\
                                 &= (\omega^{i x_\alpha j + i b_\alpha})_{0\leq i \leq N-1, 1 \leq j \leq T} \\
                                 &= (\omega^{i x_\alpha j} \omega^{i b_\alpha})_{0\leq i \leq N-1, 1 \leq j \leq T} \\
                                 &= ((\omega^{i x_\alpha})^j \omega^{i b_\alpha})_{0\leq i \leq N-1, 1 \leq j \leq T} \\
                                 &= \bfD\left((\omega^{i b_\alpha})_{0\leq i \leq N-1}\right) \cdot \\ 
                                 &\qquad\left((\omega^{ix_\alpha})^j\right)_{0\leq i \leq N-1, 1 \leq j \leq T},
  \end{align*} 
  where $\bfD\left((\omega^{i b_\alpha})_{0\leq i \leq N-1}\right)$ denotes the $N \times N$ diagonal matrix with diagonal entries $\omega^{0}, \omega^{b_\alpha}, \dots, \omega^{N-1}$.
  Thus, any $T\times T$ submatrix of $\bfV(\boldsymbol{\rho}, \as)$ can be expressed as the product of a diagonal matrix with non-zero diagonal entries and the transpose of a Vandermonde matrix. 
Finally, we point out that the $\omega^{ix_\alpha}, i \in \setint{0}{N-1}$ are distinct, since $x_\alpha \perp q$ and $q\geq N$. 
 The proof for $\bfV(\boldsymbol{\rho}, \bs)$ follows analogous steps.
\end{proof}

\subsection{Proof of \label{app:lemmaequivconstraints}}

\lemmaequivconstraints*
\begin{proof}
  We reformulate conditions II) and III) for \cref{constr:cat}.
  \begin{enumerate}
    \item[II)] The condition is equivalent to 
        \begin{align*}
          &\nexists j_1, j_2\in \setint{0}{K-1}, i_1, i_2 \in \setint{0}{L-1}:\\
          &\quad \quad (i_1 \neq i_2\text{ or }j_1 \neq j_2) \\ 
          &\quad \quad \text{ and } (j_1-j_2)y \equiv (i_2-i_1)x \pmod q.
      \end{align*}
      Since $x\perp q$ and $y\perp q$ the condition is only violated if $i_1 \neq i_2$ \emph{and} $j_1 \neq j_2$. Without loss of generality let $i_2<i_1$.
      Then, we have \begin{align*}
          &\nexists j\in \setint{-K+1}{K-1}, i \in \setint{-L+1}{-1}: \\
          &\quad \quad jy \equiv ix \pmod q.
      \end{align*}
    \item[III)] a) The condition is equivalent to 
        \begin{align*}
            & \nexists j_1\in \setint{0}{K-1}, i_1\in \setint{0}{L-1}, \\ 
            & \quad j_2 \in \setint{0}{K-1}, j_3 \in \setint{0}{T-1}: \\
            & \quad \quad (j_1-j_2-j_3)y \equiv (-i_1-1)x \pmod q,%
        \end{align*}
        which in turn is equivalent to
        \begin{align*}
              &\nexists j\in \setint{-K-T+2}{K-1}, i \in \setint{-L}{-1}:\\
              &\quad \quad jy \equiv ix \pmod q.
      \end{align*} \\
      b) The condition is equivalent to
      \begin{align*} 
        & \nexists j_1\in \setint{0}{K-1}, i_1\in \setint{0}{L-1},\\& i_2 \in \setint{0}{T-1}, i_3 \in \setint{0}{L-1}: \\
                     & \quad \quad (j_1-K)y \equiv (i_2+i_3-i_1)x \pmod q,  %
        \end{align*}
        which in turn is equivalent to
        \begin{align*}
          &\nexists j\in \setint{-K}{-1}, i \in \setint{-L+1}{T+L-2}: \\
          & \quad \quad jy \equiv ix \pmod q.
      \end{align*} \\
      c) The condition is equivalent to \begin{align*}
          &\nexists j_1\in \setint{0}{K-1}, i_1\in \setint{0}{L-1}, \\ 
          & \quad i_2 \in \setint{0}{T-1}, j_2 \in \setint{0}{T-1}: \\
          &\quad \quad (j_1-j_2-K)y \equiv (i_2-i_1-1)x \pmod q,%
        \end{align*}
        which in turn is equivalent to
        \begin{align*}
            & \nexists j\in \setint{-K-T+1}{-1}, i \in \setint{-L}{T-2}: \\ & 
            \quad \quad jy \equiv ix \pmod q.
      \end{align*}
  \end{enumerate}

  To summarize, II) and III) are fulfilled if $jy \not\equiv ix \pmod q$
      for all pairs $i, j$ given by \begin{itemize}
        \item $j\in \{\setint{-K+1}{K-1}\}\setminus \{0\}, i \in \setint{-L+1}{-1}$,
        \item $j\in \setint{-K-T+2}{K-1}, i \in \setint{-L}{-1}$, 
        \item $j\in \setint{-K}{-1}, i \in \setint{-L+1}{T+L-2}$, and
        \item $j\in \setint{-K-T+1}{-1}, i \in \setint{-L}{T-2}$.
      \end{itemize}
      Note that the second pair of intervals contains the first.
\end{proof}

\subsection{Proof of \cref{lemma:latticesolutions}} \label{app:lemmalatticesolutions}

\lemmalatticesolutions*
\begin{proof}
  We analyze the set of solutions to the equivalent equation\begin{align}
    ix - jy \equiv 0 \pmod q. \label{eq:latticeequation}
  \end{align}
  By the definition of $y$ in~\cref{constr:cat}, $\sol1 \defeq (i=-\Tm, j=\Ks)$ is a solution to \eqref{eq:latticeequation}.
  Another solution is given by $\sol2 \defeq (i=\Ls, j=\Tm)$, since  
  \begin{align*}
    x\Ls - y\Tm \bmod q &= \frac{\Ls}{\Tm} \left(\frac{\Tm}{\Ls}x\Ls - \frac{\Tm}{\Ls} y\Tm \right) \bmod q \\
                        &= \frac{\Ls}{\Tm} \left(x\Tm + y\frac{-\Tm^2}{\Ls} \right) \bmod q \\
                        &= \frac{\Ls}{\Tm} \left(x\Tm + y\frac{q-\Tm^2}{\Ls} \right) \bmod q \\
                        &= \frac{\Ls}{\Tm} \left(x\Tm + y\Ks \right) \bmod q \\
                        &= 0.
  \end{align*}
  Integer linear combinations of solutions to \eqref{eq:latticeequation} are also solutions, i.e. \sol1 and \sol2 span a lattice $\lat \subset \bbZ^2$ of solutions to \eqref{eq:latticeequation}.
  The set of \emph{all} solutions to \eqref{eq:latticeequation} forms a lattice $\lat^\prime$ s.t. $\lat \subseteq \lat^\prime \subset \bbZ^2$.

  Finally, we show that $\lat$ is identical to $\lat^\prime$ rather than a sublattice, by analyzing the density of points in $\lat$ and $\lat^\prime$.

  The lattice $\lat^\prime$ has a density of $1/q$ points per unit area. This can be seen, for example, by analyzing the probability of $ix\equiv jy \pmod q$ for $i$ and $j$ uniformly distributed over $\setint{0}{q-1}$ and using the fact that $\lat^\prime$ is periodic with period $q$ in the $i$ and $j$ direction. 
  We can confirm that the lattice \lat indeed has the same density by computing the area of its fundamental parallelogram.
\end{proof}

\subsection{Proof of \cref{lemma:boundingbox}}
\lemmaboundingbox*

\begin{proof}
  According to \cref{lemma:latticesolutions}, the lattice $\lat$ spanned by $\sol1 = (i=-\Tm, j=\Ks)$ and $\sol2 = (i=\Ls, j=\Tm)$ contains all solutions to $ix\equiv jy \pmod q$.
  The following proof is by a geometric argument in the $(i,j)$-plane. Let $\rect$ denote the rectangle defined by  $-L \leq i \leq T+L-1$ and $-K-T+1 \leq j \leq K-1$.
  We partition the set of points in $\lat$ and treat the parts separately:
  \begin{enumerate}[label=\Alph*)]
    \item $\{0\sol1+0\sol2, -\sol1+0\sol2, 0\sol1+\sol2\}$: There is nothing to show, as these are the exceptions listed in the lemma.
    \item $\{a \sol1 + b\sol2 \given a\geq 1, b\geq 0\}$: These points are above \rect since $j=a\Ks+b\Tm>K-1$ for $a\geq 1, b\geq 0$.
    \item $\{a \sol1 + b\sol2 \given a\leq -1, b\geq 1\}$: These points are right of \rect since $i=-a\Tm+b\Ls>T-1+L$ for $a\leq -1, b\geq 1$.
    \item $\{a \sol1 + b\sol2 \given a\geq 0, b\leq -1\}$: These points are left of \rect since $i=-a\Tm+b\Ls<-L$ for $a\geq 0, b\leq -1$.
    \item $\{a \sol1 + b\sol2 \given a\leq -1, b\leq -1\}$: These points are below \rect since $j= a\Ks+b\Tm < -K-T+1$ for $a\leq -1, b\leq -1$.
    \item $\{a\sol1 + b\sol2 \given a=0, b\geq 2\}$: These points are right of \rect since $i=b\Ls > T+L-1$ for $b\geq 2$.
    \item $\{a\sol1 + 0\sol2 \given a\leq -2\}$: These points are below \rect since $j=a\Ks < -K-T+1$ for $a\leq -2$.
  \end{enumerate}
  The following table shows that all points $a\sol1 + b\sol2,~a,b\in \bbZ$ are included in one of the cases above:
  \vspace{1em}
  
  \centering\begin{tabular}{c|cccc}
              & $b\leq -1$ & $b=0$ & $b=1$ & $b\geq2$ \\ \hline
    $a\geq1$  & D)          & B)     & B)     & B)       \\
    $a=0$     & D)          & A)     & A)     & F)       \\
    $a=-1$    & E)          & A)     & C)     & C)       \\
    $a\leq-2$ & E)          & G)     & C)     & C)      
    \end{tabular}

\end{proof}

\subsection{Proof of \cref{lemma:intersections}}
\lemmaintersections*
\begin{proof}
  First, we point out $2\Tm-\kappa$ elements in the intersection $\TR \cap \BR$, then we show that no others exist.
  We have\begin{align*}
    \TR &= -x + \setint{0}{K+T-2}y \bmod q \\
    \BR &= \setint{0}{T-1}x+Ky + \setint{0}{T-1}y-x \bmod q \\
        &= \setint{-1}{T-2}x + \setint{K}{K+T-1}y \bmod q,
  \end{align*}
  and let
  \begin{align*}
    \BR^{(1)} &\defeq \{-x, (T-2)x\} + \setint{K}{K+T-1}y \bmod q \\
    \BR^{(2)} &\defeq \setint{0}{T-3}x + \setint{K}{K+T-1}y \bmod q.
  \end{align*}
  The numbers in $\cI^{(1)} \defeq -x+\setint{K}{K+T-2}y \bmod q$ and $\cI^{(2)} \defeq -x+\setint{0}{T-\kappa-2}y \bmod q = (T-2)x + \setint{K+1+\kappa}{K+T-1}y \bmod q$ occur in both $\TR$ and $\BR^{(1)}$. 
  We have $|\cI^{(1)}\cup\cI^{(2)}|=2\Tm-\kappa$, since $y\perp q$ and $K+T-1<q$.
  
  Let\begin{align*}
    \BR_\mathrm{leftover}^{(1)} \defeq& \BR^{(1)} \setminus (\cI^{(1)} \cup \cI^{(2)}) \\
                                      =& (T-2)x+\setint{K}{K+\kappa}y \\
                                      &\cup \{-x+(K+T-1)y\}\bmod q \\
                                      =& -x-\setint{1}{1+\kappa}y \\
                                      &\cup \{-x+(K+T-1)y\}\bmod q.
  \end{align*}
  By comparing to the expression of $\TR$, we can see that $\BR_\mathrm{leftover}^{(1)} \cap \TR = \emptyset$.

  It remains to show that $\BR^{(2)} \cap \TR = \emptyset$.
  If not, then there exist integers $0\leq i_1\leq T-3, K \leq j_1 \leq K+T-1, 0\leq j_2 \leq K+T-2$ s.t.
  \begin{align*}
    i_1x + j_1 y \equiv -x + j_2y \pmod q,
  \end{align*}
  i.e. there exist integers $1 \leq i \leq T-2$ and $-K-T+1 \leq j \leq T-2$, s.t. $ix \equiv jy \pmod q$. This would contradict \cref{lemma:boundingbox}.

  The equality $|\BL \cap \BR| = 2\Tm - \lambda$ follows from analogous arguments.
\end{proof}

\subsection{Proof of \cref{lemma:gasprsvalid}} \label{app:lemmagasprsvalid}

\lemmagasprsvalid*
\begin{proof}
Let $\TL = \{\ap\} + \{\bp\}$, $\TR = \{\ap\} + \{\bs\}$, $\BL = \{\as\} + \{\bp\}$, $\BR = \{\as\} + \{\bs\}$, and $\boldsymbol{\gamma}=\vect(\{\ap||\as\}+\{\bp||\bs\})$.

  We show that the conditions hold separately:
  Condition I) holds trivially.
  Condition II) requires that $|\TL|= KL$ and holds since $\TL=\setint{0}{KL-1}$. 
  Condition III) requires that $\TL$ does not intersect either of $\TR$, $\BL$, and $\BR$, which holds as all elements of $\TL$ are strictly smaller than the elements of $\TR$, $\BL$, and $\BR$. 
  Condition IV) states, that the vectors $\ap || \as$ and $\bp || \bs$ (individually) must not contain duplicate entries, which holds since they are strictly increasing. 
\end{proof}

\subsection{Proof of \cref{lemma:dogrsvalid}} \label{app:lemmadogrsvalid}

\lemmadogrsvalid*
\begin{proof}
  Let $\TL = \{\ap\} + \{\bp\}$, $\TR = \{\ap\} + \{\bs\}$, $\BL = \{\as\} + \{\bp\}$, $\BR = \{\as\} + \{\bs\}$, and $\boldsymbol{\gamma}=\vect(\{\ap||\as\}+\{\bp||\bs\})$.

  We show that the conditions hold separately:
  Condition I) holds trivially.
  Condition II) requires that $|\TL|= KL$. We have $\TL = \{i (K+r) + j | i\in \setint{0}{L-1}, j \in \setint{0}{K-1}\}$. Long division of $i(K+r) + j$ by $K+r$ uniquely recovers $i$ and $j$, i.e. there are $KL$ distinct elements.
  Condition III) requires $\TL \cap \TR = \TL \cap \BL = \TL \cap \BR = \emptyset$.  Since, all elements of $\bs$ are larger than all elements of $\TL$, we have $\TL \cap \TR = \TL \cap \BR = \emptyset$. All elements of $\TR$ are congruent to a number in $\setint{0}{K-1}$ modulo $K+r$, while all elements of $\BL$ are congruent to a number in $\setint{K}{K+r-1}$ modulo $K+r$, i.e. $\TL \cap \BL = \emptyset$.

  Condition IV) states, that the vectors $\ap || \as$ and $\bp || \bs$ (individually) must not contain duplicate entries, which holds since they are strictly increasing. 
\end{proof}

\vspace{10em} %

\subsection{Supplementary Figures} \label{app:supplementaryfigures}

\begin{figure}[htb]
\centering
\begin{subfigure}{0.24\textwidth}
    \centering
    \resizebox{.9\textwidth}{!}{
    \input{tikz/CATx1_444}}
    \caption{\small $x=1, y=13$.}
    \label{fig:CATx1_444}
\end{subfigure}
\hfill
\begin{subfigure}{0.24\textwidth}
    \centering
    \resizebox{.9\textwidth}{!}{
    \input{tikz/CATx5_444}}
    \caption{\small $x=5, y=13$.}
    \label{fig:CATx5_444}
\end{subfigure}
\hfill
\begin{subfigure}{0.24\textwidth}
    \centering
    \resizebox{.9\textwidth}{!}{
    \input{tikz/CATx13_444}}
    \caption{\small $x=13, y=33$.}
    \label{fig:CATx13_444}
\end{subfigure}
\hfill
\begin{subfigure}{0.24\textwidth}
    \centering
    \resizebox{.9\textwidth}{!}{
    \input{tikz/CATx21_444}}
    \caption{\small $x=21, y=1$.}
    \label{fig:CATx11_444}
\end{subfigure}
\caption{\small CATs following \cref{constr:cat} for $K=L=T=4$ with different $x$ and $y$. All have $N=q=34$. For these parameters, $\kappa=\lambda=0$.}
\label{fig:CATxs444}
\end{figure}

\begin{figure}[htb]
\centering
    \centering\resizebox{0.9\linewidth}{!}{
    \input{tikz/GASPrs_smin.tex}}
  \caption{\small The $s$ that minimizes $\ngasprs(K,L,T)$. The value of $r$ is optimized for each $s$. Whenever the displayed value does not equal $T$, \gasprs outperforms \gaspr.}

  \label{fig:GASPrs_best_s_heatmap}
\end{figure}

\begin{figure*}[htb]
\centering
\begin{subfigure}{0.49\linewidth}
    \centering\resizebox{0.7\linewidth}{!}{
    \input{tikz/SPArs776.tex}}
    \caption{\small DOG$_{r=1,s=3}$ with $N=88$.}
    \label{fig:SPArs776}
\end{subfigure}
\hfill
\begin{subfigure}{0.49\linewidth}
    \centering\resizebox{0.7\linewidth}{!}{
    \input{tikz/GASPrs776.tex}}
    \caption{\small GASP$_{r=2,s=3}$ with $N=89$.}
    \label{fig:GASPrs776}
\end{subfigure}
\hfill
\begin{subfigure}{0.49\linewidth}
    \centering\resizebox{0.7\linewidth}{!}{
    \input{tikz/CATx8_776.tex}}
    \caption{\small CAT$_{x=8}$ with $N=89$.}
    \label{fig:CATx776}
\end{subfigure}
\hfill
\begin{subfigure}{0.49\linewidth}
    \centering\resizebox{0.7\linewidth}{!}{
      \input{tikz/GASPr776.tex}}
      \caption{\small GASP$_{r=3}$ with $N=91$.}
    \label{fig:GASPr776}
\end{subfigure}
\caption{\small The best \dogrs, \gasprs, and \gaspr as well as a \catx for $K=L=7$ and $T=6$.} %
\label{fig:GASP_v_GASPrs}
\end{figure*}

\begin{figure*}[t!]
  \begin{subfigure}{0.32\linewidth}
  \centering\resizebox{\linewidth}{!}{\input{tikz/bestscheme_T=4}}
    \caption{\small $T=4$.}
    \label{fig:bestscheme_T4}
\end{subfigure}
\hfill
\begin{subfigure}{0.32\linewidth}
  \centering\resizebox{\linewidth}{!}{\input{tikz/bestscheme_T=8}}
    \caption{\small $T=8$.}
    \label{fig:bestscheme_T8}
\end{subfigure}
\hfill
\begin{subfigure}{0.32\linewidth}
  \centering\resizebox{\linewidth}{!}{\input{tikz/bestscheme_T=20}}
    \caption{\small $T=20.$}
    \label{fig:bestscheme_T20}
\end{subfigure}
\caption{\small The color indicates which scheme uses the lowest number of workers for the given values of $K\geq L$, and $T$. For $L>K$ the problem can be transposed as $\bfA \bfB = (\bfB^T \bfA^T)^T$. The numbers indicate how many workers the best scheme saves over the second best. (When multiple schemes are tied, i.e. when the number shown is zero, the scheme listed higher in the legend determines the color of the cell.)}
\label{fig:bestschemes_KneqL}
\end{figure*}

\begin{figure*}
  \centering\resizebox{0.8\linewidth}{!}{
  \input{tikz/kappa_lambda_big.tex}}
  \caption{\small The value of $\kappa$ for a given $K$ and $T$, or equivalently, the value of $\lambda$ for a given $L$ and $T$. The smallest non-negative integer s.t. $(K+1+\kappa) \perp (T-1)$ and $(L+1+\lambda) \perp (T-1)$. The relevant area of $T\leq K, L$ is shown. Each column is periodic with a period equal to the square-free kernel of $T-1$. (The product of the distinct prime factors of $T-1$).}
  \label{fig:kappalambda}
\end{figure*}

\end{document}